\documentclass[acmsmall,screen]{acmart}

\usepackage{mathtools}
\usepackage{tikz}
\usepackage{listings}
\usepackage{mathpartir}
\usepackage{pl-syntax}
\usepackage{relsize}
\usepackage{multirow}

\newcommand{\code}[1]{\lstinline{#1}}
\newcommand{\nat}{\mathbb{N}}
\newcommand{\rat}{\mathbb{Q}}
\newcommand{\type}{\mathsf{t}}
\newcommand{\func}{\mathsf{f}}
\newcommand{\dist}{\mathsf{d}}
\newcommand{\one}{\mathsf{1}}
\newcommand{\Bool}{\mathsf{Bool}}
\newcommand{\true}{\mathsf{true}}
\newcommand{\false}{\mathsf{false}}
\newcommand{\fst}{\mathsf{fst}}
\newcommand{\snd}{\mathsf{snd}}
\newcommand{\Var}[2]{\mathsf{var}(#1 : #2)}
\newcommand{\App}[4]{\mathsf{app}_{#2 \to #3} \ #1 \ #4}
\newcommand{\ret}[1]{\mathsf{ret} \ #1}
\newcommand{\Samp}[4]{\mathsf{samp}_{#2 \to #3} \ #1 \ #4}
\newcommand{\samp}[2]{\mathsf{samp} \ #1 \ #2}
\newcommand{\Read}[2]{\mathsf{read}(#1 : #2)}
\renewcommand{\read}[1]{\mathsf{read} \ #1}
\newcommand{\ifte}[3]{\mathsf{if} \ #1 \ \mathsf{then} \ #2 \ \mathsf{else} \ #3}
\newcommand{\zero}{\mathsf{0}}
\newcommand{\assign}[2]{#1 \coloneqq #2}
\newcommand{\Par}[2]{#1 \; \| \; #2}
\newcommand{\new}[3]{\mathsf{new} \ #1 : #2 \ \mathsf{in} \ #3}
\newcommand{\approxcong}[6]{#1 \cong #2 : #3 \to #4 \ \context \ #6 \ \mathsf{for \ axiom} \ \# \ #5}
\renewcommand{\approxeq}[6]{#1 \approx #2 : #3 \to #4 \ \cnt \ #5 \ \context \ #6}
\renewcommand{\int}[1]{\llbracket #1 \rrbracket}
\newcommand{\val}[1]{\mathsf{val} \ #1}
\newcommand{\outstep}[2]{\xmapsto{#1 \, \coloneqq \, #2}}
\newcommand{\eval}[1]{{#1 \hspace{-2.5pt} \Downarrow}}
\newcommand{\valueat}[2]{#1|_{\mathsf{val}(#2)}}
\renewcommand{\max}{\mathsf{max}}
\newcommand{\tmnorm}[1]{\Vert #1 \Vert}
\newcommand{\TM}{\mathsf{TM}}
\newcommand{\Adv}{\mathsf{Adv}}
\newcommand{\round}{\mathsf{rnd}}
\newcommand{\tape}{\mathsf{tape}}

\newcommand{\err}{\mathsf{err}}
\newcommand{\sem}{\mathsf{sem}}
\newcommand{\adv}{\mathsf{adv}}
\newcommand{\cnt}{\mathsf{count}}

\newcommand{\context}{\mathsf{ctxt}}

\newcommand{\poly}{\mathsf{poly}}
\newcommand{\lapproxcong}[7]{#1 \cong_{#7} #2 : #3 \to #4 \ \context \ #6 \ \mathsf{for \ axiom} \ \# \ #5}
\newcommand{\lapproxeq}[7]{#1 \approx_{#7} #2 : #3 \to #4 \ \cnt \ #5 \ \context \ #6}
\newcommand{\interaction}[3]{#1 \xrightleftharpoons{#3} #2}
\newcommand{\tmenc}[1]{\mathsf{Enc}[#1]}
\newcommand{\QAnnRead}[3]{\mathsf{read}[a](#1 : #2)}
\newcommand{\parm}{\mathord{\color{black!33}\bullet}}%

\newtheorem{lemma}{Lemma}
\newtheorem*{lemma*}{Lemma}
\newtheorem{theorem}{Theorem}
\newtheorem*{theorem*}{Theorem}
\newtheorem{definition}{Definition}



\begin{document}

\title{Concrete Security Bounds for Simulation-Based Proofs of Multi-Party Computation Protocols}

\author{Kristina Sojakova}
\email{k.sojakova@vu.nl}
\orcid{0000-0003-4880-1416}
\affiliation{
 \institution{Vrije Universiteit Amsterdam}
 \country{Netherlands}
} 

\author{Mihai Codescu}
\email{mscodescu@gmail.com}
\orcid{0000-0002-7702-8955}
\affiliation{
 \institution{Research Group of the NLNet Project IPDL}
 \country{Romania}
}

\author{Joshua Gancher}
\email{j.gancher@northeastern.edu}
\orcid{0000-0003-2257-7073}
\affiliation{
 \institution{Northeastern University Boston}
 \country{United States of America}
}

\begin{abstract}
The \emph{concrete security} paradigm aims to give precise bounds on the probability that an adversary can subvert a cryptographic mechanism. This is in contrast to asymptotic security, where the probability of subversion may be \emph{eventually} small, but
large enough in practice to be insecure. Fully satisfactory concrete security bounds for Multi-Party Computation (MPC) protocols are difficult to attain, as they require reasoning about the \emph{running time} of cryptographic adversaries and reductions.

In this paper we close this gap by introducing a new foundational approach that allows us to automatically compute concrete security bounds for MPC protocols. We take inspiration from the meta-theory of IPDL, a prior approach for formally verified distributed cryptography, to support reasoning about the runtime of protocols and adversarial advantage. For practical proof developments, we implement our approach in Maude, an extensible logic for equational rewriting.

We carry out four case studies of concrete security for simulation-based proofs. Most notably, we deliver the first formal verification of the GMW MPC protocol over $N$ parties. To our knowledge, this is the first time that formally verified concrete security bounds are computed for a proof of an MPC protocol in the style of Universal Composability. Our tool provides a layer of abstraction that allows the user to write proofs at a high level, which drastically simplifies the proof size. For comparison, a case study that in prior works required 2019 LoC only takes 567 LoC, thus reducing proof size by 72\%.
\end{abstract}

\maketitle

\section{Introduction}
Advanced distributed cryptographic protocols such as Multi-Party
Computation (MPC) have the potential to enable new, privacy-preserving modes of computing. However, their inherent complexity introduces new risks, including the risk that the protocol itself (or protocol optimizations employed by implementations) is insecure. The commonly-used security definition for MPC is simulation-based security in the style of Universal Composability, or UC~\cite{uc}. UC provides strong security guarantees that are robust under an embedding of the protocol into a larger distributed system. However, UC-style proofs for MPC protocols are generally very difficult, as they require complex bisimulation-based security arguments in conjunction with low-level runtime analysis.

While a number of prior approaches for verifying MPC protocols have been proposed~\cite{mpc1,mpc2,mpc3,mpc4,ipdl}, most of these approaches deliver standalone security definitions not compatible with
UC~\cite{mpc1,mpc2,mpc3,mpc4}, which harms composability. In contrast, the IPDL system~\cite{ipdl} can deliver UC-style security results for MPC protocols via a convenient \emph{equational} style of reasoning. Prior work has shown that IPDL can reason about realistic protocols, including a two-party variant of the 
GMW~\cite{gmw} MPC protocol, along with various Oblivious
Transfer~\cite{beaver1995precomputing} protocols. 

However, a number of security-critical caveats remain. None of the prior approaches for verifying MPC adequately reason about the \emph{runtime} of adversaries and simulators constructed during the
proof. Reasoning about runtime is essential for security, since nearly all cryptographic proofs contain \emph{reductions} of the form ``the probability that $A$ breaks protocol $P$ is bounded by the probability that the reduced adversary $R(A)$ breaks indistinguishability assumption $Q$``. Here $R(A)$ is an adversary whose interaction with $Q$ mimics the interaction of the original adversary $A$ with $P$. If $R(A)$'s runtime is not adequately bounded, then the probability that $R(A)$ breaks the assumption $Q$ may be $1$, rendering the security result essentially meaningless. This problem is even more pronounced for large protocols such as MPC, since the constructed simulators also become increasingly complex.

In this work, we address this issue for MPC protocols. Our method supports \emph{concrete security}~\cite{bellare1997concrete}, which gives precise bounds on the probability $\epsilon(t)$ that an attacker running in time $t$ violates the security guarantees of the system. Our novel strategy to obtain these bounds is to bound the size of the IPDL program context, and analyze the runtime of a Turing Machine that interprets IPDL programs. This strategy allows us to obtain practical bounds for MPC: indeed, \emph{to our knowledge, there is no other formally verified proof of an MPC protocol that reasons about concrete security and runtime of adversaries/simulators}.

To enable fast and scalable formal proofs, we implemented our proof system in the equational rewrite tool Maude~\cite{maude} as an extension of
SpeX~\cite{spex} and equipped it with a Domain-Specific Language (DSL) for concise proofs that automatically compile down to lower-level Maude code. Using our implementation, we carry out four different case studies. To highlight the scalability of our approach, we present a new, fully mechanized proof of \emph{simulation-based security} for the \emph{Multi-Party GMW} MPC protocol defined over an arbitrary Boolean circuit and for arbitrarily many parties. Crucially, our tool \emph{automates} the computation of concrete security bounds, which would otherwise be infeasible to carry out for a protocol of this size.

\subsection{Contributions}
In this work, we develop:

\begin{itemize}
\item a proof system that automatically computes concrete security bounds for composable simulation-based proofs of MPC protocols;
\item a formally verified proof of the \emph{GMW} MPC protocol with $N$ parties, which to our knowledge marks the first time that formally verified concrete security bounds have been computed for a UC-style proof of an MPC protocol;
\item an accompanying Maude implementation that automatically computes the aforementioned bounds and a DSL for writing proofs that hides most low-level details from the user. This dramatically  simplifies the proof size: \emph{e.g.}, excluding definitions, the proof of the \emph{Coin Flip} case study in~\cite{ipdl} takes 1905 LoC, whereas we deliver it in 256 LoC. Our proofs also significantly outperform their Coq equivalents from~\cite{ipdl} in terms of runtime: \emph{e.g.}, 5 seconds vs. a few minutes for the \emph{Coin Flip} case study.
\end{itemize}

\paragraph{Limitations of our system} Since we build upon the process calculus in \cite{ipdl}, we are only able to support protocols expressible in IPDL. Specifically, IPDL only considers protocols with static communication topologies and static security. Furthermore, it does not consider protocols that exhibit threshold behavior such as consensus protocols. As IPDL is targeted towards UC security, it also does not consider proofs that use rewinding.

\paragraph{Limitations of our proof effort} Our proof of the GMW protocol assumes that party $N$ is semi-honest and party $N+1$ is honest. While this is not fully general, all other cases are either trivial (all parties honest/all parties semi-honest) or can be essentially reduced to the aforementioned case. We note that the GMW protocol is not secure against a malicious adversary.

\paragraph{Structure of paper} In Section~\ref{sec:overview} we briefly review simulation-based security, present IPDL, and illustrate our DSL on a simple example. In Section~\ref{sec:approximate}, we give the syntax and semantics of our cost-aware proof system for simulation-based security, and present our soundness results. In
Section~\ref{sec:case_studies}, we outline our proof of the \emph{$N$-Party GMW} protocol and briefly describe the other case studies. We conclude by indicating some directions for future work.

\section{Related Work}
This paper is part of a long line of formal verification efforts for MPC and similar protocols. Some works target domain-specific, standalone security notions for protocols~\cite{mpc1,mpc2,mpc3,mpc4}, while others~\cite{ipdl,easyuc,advcompl,crypthol-cc} aim to construct general frameworks for \emph{simulation-based security} in the style of UC. While runtime analysis is in a sense required for sound cryptographic reasoning, almost all of the above works (with the exception of \cite{advcompl} and \cite{ipdl}) declare reasoning about runtime out of scope, and instead defer to the reader to ensure that all relevant cryptographic reductions have a reasonable running time.

The line of work proposed in \cite{ipdl} shows that an equational proof strategy is useful for proving MPC and related protocols secure. However, in lieu of reasoning about runtime, \cite{ipdl} relies on symbolic bounds, which provide some measure of complexity for a syntactic simulation context. However, \cite{ipdl} does not give a low-level computational semantics to protocols, nor does it carry out any cryptographic reductions. It is therefore unclear what such a symbolic bound means for the runtime of adversaries and simulators. In this paper, we build upon the process calculus introduced in \cite{ipdl} with a new cost-aware proof system and semantics that reason explicitly about the runtime of cryptographic adversaries and simulators. In particular, we prove \emph{concrete} security bounds of the form $\epsilon(t)$, where $t$ is the running time of the adversary.

The system by Barbosa et al.~\cite{advcompl} extends EasyCrypt~\cite{easycrypt} by a cost-aware Hoare logic and uses it to analyze a secure channel protocol. However, this runtime analysis only applies to sequential programs, while the process calculus of \cite{ipdl} natively handles concurrency. Thus, while~\cite{advcompl} considers UC, their computational model assumes that the protocol follows a stack discipline, which is not a good fit for MPC.

Squirrel~\cite{squirrel} and CryptoVerif~\cite{cryptoverif} are two popular tools that deliver concrete security bounds for large classes of cryptographic protocols. However, the bulk of the proof for an MPC protocol consists of manipulations such as inlining the computation from one channel to another, or removing parts of the protocol that have become unused after simplification. The process calculus of \cite{ipdl} was explicitly designed around such congruences, which do not have counterparts in Squirrel or CryptoVerif. This design choice enabled us to carry out the largest formally verified UC-style proof of an MPC protocol that we are aware of.

Overture~\cite{overture} is a recently-proposed system for proving security properties of MPC protocols via checking secrecy and integrity hyperproperties. While it offers automatic proofs, it does not prove UC-style properties, nor analyze the runtime of simulators. Owl~\cite{owl} uses an information-flow type system to deliver modular proofs of protocols that use cryptographic mechanisms (rather than reason \emph{about} them, which is required for MPC). Resource-aware session types (RAST)~\cite{rast} allow one to embed runtime guarantees into session-typed protocols. While RAST and our work both target concurrent process calculi, the protocols we consider do not naturally carry session types, and instead have a low-level computational interpretation in terms of Turing Machines. In particular, it is unclear how to embed MPC into RAST, or use session types in general to perform cryptographic security proofs. GAuV \cite{gauv} uses graph transformations to automate proofs of semi-honest security for concrete instances of the BGW protocol (\emph{i.e.}, for a fixed number of parties and a fixed circuit).

\section{Overview: Simulation-Based Security and IPDL}\label{sec:overview}
\newcommand{\Ideal}{\mathsf{Ideal}}
\newcommand{\Real}{\mathsf{Real}}
\newcommand{\Sim}{\mathsf{Sim}}
\newcommand{\xor}{\mathsf{xor}}
\newcommand{\flip}{\mathsf{flip}}
\newcommand{\In}{\mathsf{In}}
\newcommand{\Key}{\mathsf{Key}}
\newcommand{\Ctxt}{\mathsf{Ctxt}}
\newcommand{\LeakCtxtAliceAdv}{\mathsf{LeakCtxt Alice\_adv}}
\newcommand{\LeakCtxtIdAdv}{\mathsf{LeakCtxt id\_adv}}

We now briefly describe simulation-based security, give an overview of the IPDL process calculus from \cite{ipdl}, and illustrate our  approach on a simple running example.

\subsection{Simulation-Based Security}
Simulation-based security relates the behavior of a protocol to that of an idealization, where cryptographic mechanisms are replaced by trusted resources secure by construction. In this sense, an idealization is a specification for the behavior that the real-world protocol aims to approximate with cryptographic methods. For example, a real-world protocol utilizes encryption to securely send a message from Alice to Bob over a public network. The idealization instead relies on a trusted third party that securely obtains the message from Alice and forwards it to Bob.

The simulation-based paradigm, as employed \emph{e.g.} in UC and Constructive Cryptography~\cite{constructive-crypto} is very powerful and unifies various other security notions such as secrecy and integrity. In UC-style security, a protocol is an interactive system consisting of \emph{parties}, \emph{e.g.} Alice and Bob, and \emph{functionalities}, \emph{e.g.} a public network that forwards messages from Alice to Bob, or a key generation mechanism that randomly generates a secret key and securely delivers it to Alice and Bob.

Formal proofs in this setting amount to showing \emph{observational
equivalences} $P \approx Q$ between a ``real'' protocol $P$, and an ``ideal''
protocol $Q$. These proofs typically take the form of a sequence of exact ($=$) and approximate $(\approx)$ equality steps:
\[ P = P_1 \approx Q_1 = P_2 \approx Q_2 = \ldots = P_n \approx Q_n = Q \]
As observed in \cite{ipdl}, a typical approximate equality step $Q_i \approx
P_i$ consists of replacing the left-hand side of an \emph{indistinguishability
assumption} $G_i \approx H_i$ by its right-hand side in a common context $R_i$; \emph{i.e.}, $P_i$ arises as $R_i[G_i]$ and $Q_i$ as $R_i[H_i]$. 
An example of such an assumption is IND-CPA, which states that encryptions of adversarially chosen messages are indistinguishable from encryptions of zeros (in the absence of a decryption oracle).

\subsection{Background: IPDL}

IPDL~\cite{ipdl} is a process calculus for distributed cryptographic protocols (\emph{e.g.}, MPC) that enables one to prove simulation-based security results similar to those analyzed in UC. IPDL \emph{protocols} are composed of mutually interacting \emph{reactions}, which are sequential monadic programs that probabilistically compute an \emph{expression}. In the context of a protocol, a reaction operates on a unique \emph{channel} and may read from other channels, thereby utilizing computations coming from other reactions. 

Aside from a formal process calculus, the original paper~\cite{ipdl} defines two \emph{equational} proof systems: an \emph{exact} equational logic for proving perfect equivalences between protocols, and an \emph{approximate} logic for proving computational indistinguishability results. The exact equational logic is proven sound in terms of \emph{bisimulations}, while the approximate logic is proven sound in terms of \emph{computational reductions}.

Following~\cite{ipdl}, we assume a user-defined \emph{signature} that specifies the base types and the (probabilistic) functions we have at our disposal:

\begin{definition}[Signature]
A signature $\Sigma$ consists of:
\begin{itemize}
\item type constants $\type$,
\item function symbols $\func : \sigma \to \tau$, and
\item distribution symbols $\dist : \sigma \to \tau$.
\end{itemize}
\end{definition}

We summarize the syntax of IPDL in Figure~\ref{fig:syntax}. Data types and expressions are standard. Here $\one$ denotes the unit type and $\checkmark$ the canonical inhabitant of $\one$. The expression $\App{\func}{\sigma}{\tau}{e}$ denotes the application of the function symbol $\func : \sigma \to \tau$ declared in the signature $\Sigma$ to an expression $e$. Similarly, the reaction $\Samp{\dist}{\sigma}{\tau}{e}$ denotes the application of the distribution symbol $\dist : \sigma \to \tau$ declared in the signature $\Sigma$ to an expression $e$.

The reaction $\Read{c}{\tau}$ denotes the read of a value of type $\tau$ from the channel $c$. We also have branching $(\ifte{e}{R_1}{R_2})$ and the standard monadic operations of return $(\ret{e})$ and bind $(x : \sigma \leftarrow R; \ S)$. At the protocol level, we have the trivial protocol $\zero$, the single-channel protocol $\assign{o}{R}$ that assigns a reaction $R$ to the channel $o$, the parallel composition $\Par{P}{Q}$ of two protocols, and the spawning $\new{o}{\tau}{P}$ of a new internal channel $o$ of type $\tau$ for use in $P$.

In our version of the IPDL syntax, references $\Var{x}{\tau}$ to variables and $\Read{c}{\tau}$ to channels include a typing annotation. We will need these later on when encoding an IPDL construct as a sequence of symbols on a Turing Machine tape; knowing the type $\tau$ will allow us to allocate the correct number of bits for the variable $x$ or the channel $c$.

\begin{figure}[ht]
\begin{syntax}

  \abstractCategory[Variables]{x, y, z}
  \abstractCategory[Channels]{i, o, c}
	
	\category[Channel Sets]{I, O}
    \alternative{\{c_1, \ldots, c_n\}}

  \category[Data Types]{\tau, \sigma}
    \alternative{\type}
		\alternative{\one}
    \alternative{\Bool}
    \alternative{\tau_1 \times \tau_2}

  \category[Expressions]{e}
    \alternative{\Var{x}{\tau}}
    \alternative{\checkmark}
	  \alternative{\true}
	  \alternative{\false}		
	  \alternative{\App{\func}{\sigma}{\tau}{e}}
	  \alternative{(e_1,e_2)}  
	  \\  
	  \alternative{\fst_{\sigma \times \tau} \ e}
		\alternative{\snd_{\sigma \times \tau} \ e}		

  \category[Reactions]{R, S}
    \alternative{\ret{e}}
    \alternative{\Samp{\dist}{\sigma}{\tau}{e}}
    \alternative{\Read{c}{\tau}}
    \\
    \alternative{\ifte{e}{R_1}{R_2}}
    \alternative{x : \sigma \leftarrow R; \ S}         

	\category[Protocols]{P, Q}
	  \alternative{\zero}	
	  \alternative{\assign{o}{R}}
	  \alternative{\Par{P}{Q}}
	  \alternative{\new{o}{\tau}{P}}
		
  \category[Type Contexts]{\Gamma}
    \alternative{\cdot}
    \alternative{\Gamma, x : \tau}

  \category[Channel Contexts]{\Delta}
    \alternative{\cdot}
    \alternative{\Delta, c : \tau}
\end{syntax}
\caption{Syntax of \textsf{IPDL}.}
\label{fig:syntax}
\end{figure}

Typing of protocols in IPDL has the form $\Delta \vdash P : I \to O$, where $\Delta$ is a channel context assigning types to channel names, and $I,O$ are disjoint sets of \emph{input} and \emph{output} channels, respectively. Each output channel in $O$ must be assigned a reaction inside $P$. The exact equational logic of IPDL is parameterized by a finite set of axioms of the form $\Delta \vdash P_1 = P_2 : I \to O$, where $\Delta \vdash P_1 : I \to O$ and $\Delta \vdash P_2 : I \to O$. We use these axioms to express example-specific functional assumptions, \emph{e.g.}, the correctness of an encryption/decryption scheme. Figure~\ref{fig:protocols_equality_strict} shows a few illustrative rules of the exact fragment of IPDL.

Rule \textsc{comp-new} allows us to pull a sub-protocol $P$ outside the scope of a channel declaration if the bound channel name does not appear in $P$. Rule \textsc{absorb} allows us to discard a sub-protocol that has become unused after simplification. Rule \textsc{subst} says that if the computation $R_1$ assigned to a channel $o_1$ is deterministic, then we may replace every occurrence of $\read{o_1}$ by $R_1$. Rule \textsc{fold-bind} states that if we only read from channel $c$ once in the context of a protocol, we can soundly replace $\read{c}$ by the computation assigned to $c$, even if this computation is probabilistic. Finally, rule \textsc{drop} allows us to drop a vacuous dependency on channel $o_1$ from channel $o_2$ if the computation $R_1$ assigned to $o_1$ does not introduce additional dependencies to $o_2$.

\begin{figure*}
\begin{mathpar}
\fbox{$\Delta \vdash P = Q : I \to O$}\\
\inferrule*[right=comp-new]{\Delta \vdash P : I \cup O_2 \to O_1 \\ \Delta, o : \tau \vdash Q : I \cup O_1 \to O_2 \cup \{o\}}{\Delta \vdash \Par{P}{\big(\new{o}{\tau}{Q}\big)} = \new{o}{\tau}{(\Par{P}{Q})} : I \to O_1 \cup O_2}\and
\inferrule*[right=absorb]{\Delta \vdash P : I \to O \\ \Delta \vdash Q : I \cup O \to \emptyset}{\Delta \vdash \Par{P}{Q} = P}\and
\inferrule*[right=fold-bind]{\Delta; \ \cdot \vdash R : I \to \sigma \\ \Delta; \ x : \sigma \vdash S : I \to \tau}{\Delta \vdash \big(\new{c}{\sigma}{\Par{\assign{o}{{\color{red} x \leftarrow \read{c};} \ S}}{{\color{red} \assign{c}{R}}}}\big) = \big(\assign{o}{{\color{red} x \leftarrow R; \ } S}\big)}\and
\inferrule*[right=subst]{\Delta; \ \cdot \vdash \big(x \leftarrow R_1; \ y \leftarrow R_1; \ \ret{(x, y)}\big) = \big(x \leftarrow R_1; \ \ret{(x, x)}\big)}{\Delta \vdash \big(\Par{\assign{o_1}{R_1}}{\assign{o_2}{{\color{red} x_1 \leftarrow \read{o_1}; \ } R_2}}\big) = \big(\Par{\assign{o_1}{R_1}}{\assign{o_2}{{\color{red} x_1 \leftarrow R_1; \ } R_2}}\big)}\and
\inferrule*[right=drop]{\Delta; \ \cdot \vdash R_1 : I \to \tau_1 \\ \Delta; \ \cdot \vdash R_2 : I \to \tau_2 \\ \Delta; \ \cdot \vdash \big({\color{red} x_1 \leftarrow R_1; \ } R_2\big) = R_2 : I \to \tau_2}{\Delta \vdash \big(\Par{\assign{o_1}{R_1}}{\assign{o_2}{{\color{red} x_1 \leftarrow \read{o_1}; \ } R_2}}\big) = \big(\Par{\assign{o_1}{R_1}}{\assign{o_2}{R_2}}\big)}
\end{mathpar}
\caption{Selected rules for exact equality of IPDL protocols.}
\label{fig:protocols_equality_strict}
\end{figure*}

IPDL protocols come with a natural operational semantics. We slightly generalize the semantics given in~\cite{ipdl} to support dynamic-length bitstrings. To this end, we use a special placeholder symbol $\parm$ and by abuse of terminology we refer to strings $v \in \{0,1,\parm\}^\star$ as bitstrings.

\begin{definition}[Interpretation]
An interpretation $\int{-}$ for a signature $\Sigma$ associates to:
\begin{itemize}
\item each type symbol $\type$ a subset $\subseteq \{0,1,\parm\}^{|\type|}$ of bitstrings of length $|\type| \geq 0$;
\item each function symbol $\func : \sigma \to \tau$ a function $\int{\func}$ from $\int{\sigma}$ to $\int{\tau}$;
\item each distribution symbol $\dist : \sigma \to \tau$ a function $\int{\dist}$ from $\int{\sigma}$ to distributions on $\int{\tau}$.
\end{itemize}
\end{definition}

\noindent In the above, we generalize the interpretation $\int{-}$ to all types in the obvious way. To handle partial computations, we follow~\cite{ipdl} and augment the syntax of IPDL protocols to contain intermediate bitstring values 
\begin{syntax}
\category[Protocols]{P, Q}
\alternative{\assign{o}{v}} \alternative{\dots}
\end{syntax}

We give semantics to IPDL protocols via two main small-step rules, see Figure~\ref{fig:protocols_semantics}, where we write $1[P]$ for the distribution with unit mass at the protocol $P$, and freely use a distribution in place of a reaction or a protocol to indicate the obvious lifting of the corresponding construct to distributions on protocols. As reactions are sequential monadic programs, they admit a straightforward small-step semantics $R \to \eta$, which we omit. Big-step operational semantics for protocols $P \Downarrow \eta$ performs output and internal steps in an arbitrary order until no more steps are possible, resulting in a unique distribution $\eta$ on protocols.

\begin{figure}[ht]
\begin{mathpar}
\fbox{$P \outstep{o}{v} Q$}\\
\inferrule*{P \outstep{o}{v} P'}{\Par{P}{Q} \outstep{o}{v} \Par{P'}{Q[\assign{\read{o}}{\val{v}}]}}\and
\inferrule*{Q \outstep{o}{v} Q'}{\Par{P}{Q}\outstep{o}{v} \Par{P[\assign{\read{o}}{\val{v}}]}{Q'}}\and
\inferrule*{ }{\big(\assign{o}{\val{v}}\big) \outstep{o}{v} \big(\assign{o}{v}\big)}\and
\inferrule*{P \outstep{o}{v} P' \\ o \neq c}{\big(\new{c}{\tau}{P}\big) \outstep{o}{v} \big(\new{c}{\tau}{P'}\big)}\\\\
\fbox{$P \to \eta$}\\
\inferrule*{R \to \eta}{\big(\assign{o}{R}\big) \to \big(\assign{o}{\eta}\big)}\and
\inferrule*{P \to \eta}{\Par{P}{Q} \to \Par{\eta}{Q}}\and
\inferrule*{Q \to \eta}{\Par{P}{Q} \to \Par{P}{\eta}}\and
\inferrule*{P \to \eta}{\big(\new{c}{\tau}{P}\big) \to \big(\new{c}{\tau}{\eta}\big)}\and
\inferrule*{P \outstep{c}{v} P'}{\big(\new{c}{\tau}{P}\big) \to 1[\new{c}{\tau}{P'}]}
\end{mathpar}
\caption{Small-step operational semantics for IPDL protocols.}
\label{fig:protocols_semantics}
\end{figure}

\subsection{Example: Authenticated-To-Secure Channel}

\newcommand{\key}{\mathsf{key}}
\newcommand{\msg}{\mathsf{msg}}
\newcommand{\ctxt}{\mathsf{ctxt}}
\newcommand{\zeros}{\mathsf{zeros}}
\newcommand{\gen}{\mathsf{gen}}
\newcommand{\enc}{\mathsf{enc}}
\newcommand{\dec}{\mathsf{dec}}
\newcommand{\id}{\mathsf{id}}
\renewcommand{\adv}{\mathsf{adv}}
\newcommand{\net}{\mathsf{net}}
\renewcommand{\In}{\mathsf{In}}
\newcommand{\Out}{\mathsf{Out}}
\renewcommand{\Key}{\mathsf{Key}}
\newcommand{\Send}{\mathsf{Send}}
\newcommand{\Recv}{\mathsf{Recv}}
\newcommand{\Enc}{\mathsf{Enc}}
\newcommand{\Dec}{\mathsf{Dec}}
\newcommand{\LeakMsgRcvd}{\mathsf{LeakMsgRcvd}}
\newcommand{\OkMsg}{\mathsf{OkMsg}}
\newcommand{\LeakCtxt}{\mathsf{LeakCtxt}}
\newcommand{\OkCtxt}{\mathsf{OkCtxt}}

We now revisit the running example of \cite{ipdl}. Alice wants to securely communicate $n$ messages to Bob using an authenticated channel which leaks all messages to the adversary. To do so, we will assume they share a pre-shared key, which enables them to encrypt and decrypt all messages. We show how to encode this protocol and its proof in our DSL. First, we declare the number of sessions as a parameter $n$ to our case study, with the intended interpretation that $n$ is a function of the security parameter $\lambda$:
\begin{lstlisting}
parameter n : nat .
\end{lstlisting}

\subsubsection{The Assumptions}
We use a version of the IND-CPA assumption, which states that encoding $n$ context-chosen messages with the same secret key is indistinguishable from encrypting zeros. For simplicity, we assume a type of messages with constant length, so that the constant \code{zeros} need not depend on the length. We express the assumption as an \emph{axiom} about the equality between two protocols:
\begin{lstlisting}
approx-assumption CPA :
  (fam In[i < n] :: msg) 
  (fam Enc[i < n] :: ctxt)
  inputs: fam In[i < n] |=
  new Key : key in
    (Key ::= samp gen_key ||
    (family Enc[i < n] ::= 
       m : msg <- read In[i] ;
       k : key <- read Key ;
       samp enc((m, k))))
   ~
  new Key : key in
    (Key ::= samp gen_key ||
    (family Enc[i < n] ::=
       m : msg <- read In[i] ;
       k : key <- read Key ;
       samp enc((zeros, k)))) .
\end{lstlisting}
The \code{CPA} axiom is parameterized by two
\emph{families} of channels: \code{In[i < n]}, for \code{n} input channels
carrying messages, and
\code{Enc[i < n]}, for \code{n} output channels carrying ciphertexts. 
The left side of the \code{CPA} assumption samples a key on channel \code{Key}
and, for each \code{i}, encrypts \code{In[i]} under the key. We do this by declaring a
\code{family} of protocols --- one for each \code{i} less than the parameter
\code{n} --- which reads from
\code{In[i]}, reads from \code{Key}, and samples from the distribution of
probabilistic encryptions under that key and message.  
We declare the
\code{Key} channel as \emph{internal} using \code{new} so that the outside
context cannot read it. The right side of the \code{CPA} assumption is similar,
but encrypts \code{zeros} (defined to be a constant) rather than the message. 
Importantly, the family \code{Enc[i < n]} on the right hand side still reads from
\code{In[i]}, since the two protocols must have the same (logical) \emph{timing} behaviors
between the channels.

In addition to the \code{CPA} assumption, we also have the assumption about the encryption scheme's \emph{correctness}: encrypting and decrypting must return the same message. We encode this in our DSL as an assumption similar to \code{CPA}, but since we assume that the encryption scheme is \emph{perfectly} correct, with zero probability of error, we use the declaration \code{protocol-assumption} rather than
\code{approx-assumption}:

\begin{lstlisting}
protocol-assumption enc-dec-correctness :
  (chn In :: msg) (chn Key :: key)
  (chn Enc :: ctxt) (chn Dec :: msg)
  inputs: chn In, chn Key |=
  (Enc ::= m : msg <- read In;
            k : key <- read Key;
            samp enc((m, k))) ||
  (Dec ::= c : ctxt <- read Enc;
            k : key <- read Key;
            return dec((c, k)))
   =
  (Enc ::= m : msg <- read In;
            k : key <- read Key;
            samp enc((m, k))) ||
  (Dec ::= i : msg <- read In;
            return i).
\end{lstlisting}

\subsubsection{The Protocol}
The Authenticated-To-Secure Channel protocol \code{Real} now takes the following form:
\begin{lstlisting}
protocol Real =
  new Key : key in
  newfamily Recv[i < n] : ctxt in
  newfamily Send[i < n] : ctxt in
   (Keygen || Alice || Channel || Bob)
    where Alice =
        (family Send[i < n] ::=
           m : msg <- read In[i] ;
           k : key <- read Key ;
           samp enc((m, k)))
    and Bob =
        (family Out[i < n] ::=
           c : ctxt <- read Recv[i] ;
           k : key <- read Key ;
           return dec((c, k)))
    and Channel = 
        (family Leak[i < n] ::= read Send[i]) ||
        (family Recv[i < n] ::= 
           c : ctxt <- read Send[i] ; 
           ok : unit <- read Ok[i]  ; 
           return c)
    and Keygen =
        (Key ::= samp gen_key) .
\end{lstlisting}
The body of the protocol is a parallel composition of the key generating
functionality \code{Keygen}, the two parties \code{Alice} and \code{Bob}, and
the authenticated channel functionality \code{Channel}. Alice encrypts each
input with the shared key stored on the internal channel \code{Key}, samples a
ciphertext from the resulting distribution, and sends the result to the
authenticated channel functionality on the channel \code{Send[i]}. Bob reads the
ciphertext forwarded to him from the authenticated channel functionality on the
channel \code{Recv[i]}, decrypts it with the shared key, and outputs the
plaintext on the channel \code{Out[i]}. The channels \code{Send[i]} and
\code{Recv[i]} are connected by the \code{Channel} functionality, which allows
the adversary to read/schedule messages via \code{Leak[i]} and \code{Ok[i]}.

\paragraph{Proving Protocols Secure}
Our protocol is named \code{Real} because we will compare it to an \emph{ideal} version, where Alice and Bob communicate directly  through a secure channel, without the need for encryption. As in UC, we do this by proving that the \code{Real} protocol is
indistinguishable from the \code{Ideal} protocol composed with a \emph{simulator} \code{Sim} that can emulate the \code{Leak[i]} messages without knowledge of the secret messages:
\[ \text{\code{Real}} ~= \text{\code{Ideal || Sim}}. \]
The key advantage of IPDL~\cite{ipdl} is that it enables proofs through
\emph{equational reasoning} principles. Using basic identities of protocols (\emph{e.g.}, inlining definitions of channels into other channels) and assumptions (\emph{e.g.}, \code{CPA}), one proves a
protocol secure by progressively rewriting the real protocol into its
idealization (plus the simulator).

We illustrate some key steps of the proof. The ideal functionality for our secure channel example has two output channels per session $i$: the adversarial output channel \code{LeakMsgRcvdIdAdv[i]} reads the message from \code{In[i]} and lets the adversary know that a message has been received -- by passing a term of the \code{unit} type -- but divulges nothing about the value of the message. The channel \code{Out[i]} first waits to receive a confirmation on the adversarial input channel \code{OkMsgAdvId[i]} that gives the green light to the functionality to process the message. It subsequently reads the message from \code{In[i]} on behalf of Alice, and outputs it on behalf of Bob:
\begin{lstlisting}
Ideal =
  (family LeakMsgRcvdIdAdv[i < n] ::=
     m : msg <- read In[i];
     return ()) ||
  (family Out[i < n] ::=
     okMsg : unit <- read OkMsgAdvId[i];
     m : msg <- read In[i];
     return m)
\end{lstlisting}
The simulator turns the adversarial interface of the real protocol into the adversarial interface of the ideal functionality, thereby converting any adversary for the real protocol into an adversary for the functionality. In our example, the channels \code{LeakMsgRcvdIdAdv[i]} and \code{OkCtxtAdvNet[i]} are the inputs to the simulator, while the channels \code{LeakCtxtNetAdv[i]} and \code{OkMsgAdvId[i]} are the outputs.

Hence, upon receiving the information from the ideal functionality that a message has been received, the simulator must conjure up a ciphertext to leak to the adversary. This is accomplished by randomly generating a secret key and encrypting the chosen message \code{zeros} in each session. Upon receiving the approval from the adversary for the generated ciphertext, the simulator gives the approval to the ideal functionality to output the message:
\begin{lstlisting}
Sim =
  new Key : key in
    (Key ::= samp gen_key(())) ||
    (family LeakCtxtNetAdv[i < n] ::=
       x : unit <- read LeakMsgRcvdIdAdv[i];
       k : key <- read Key;
       samp enc((zeros(()), k))) ||
    (family OkMsgAdvId[i < n] ::=
       okCtxt : unit <- read OkCtxtAdvNet[i];
       return okCtxt)
\end{lstlisting}
As the channel \code{OkMsgAdvId[i]} carries a deterministic computation, we want to inline the computation into \code{Out[i]} to yield the following:
\begin{lstlisting}
family Out[i] i < n ::=
  okCtxt : unit <- read OkCtxtAdvNet[i];
  m : msg <- read In[i];
  return m
\end{lstlisting}

This form of substitution is justified by the rule \textsc{subst} from Figure~\ref{fig:protocols_equality_strict}, which says that if the computation $R_1$ assigned to a channel $o_1$ is deterministic, then we may replace every occurrence of $\read{o_1}$ by $R_1$. However, for the rule \textsc{subst} to apply, we must first massage the protocol into a form where \code{OkMsgAdvId[i]} appears immediately next to \code{Out[i]}. In the implementation of \cite{ipdl}, this required tedious manual transformations that, \emph{e.g.}, permute channels inside a parallel composition or move channels in and out of scope of a new channel declaration (lines 87--104 of MultiChan.v). In our tool, all the necessary massaging is performed automatically, and we can simply write:
\begin{lstlisting}
subst fam OkMsgAdvId into fam Out then
subst fam LeakMsgRcvdIdAdv into fam LeakCtxtNetAdv then
absorb fam LeakMsgRcvdIdAdv then
absorb fam OkMsgAdvId
\end{lstlisting}
A crucial step in simplifying the real protocol is to conceptually separate the encryption and decryption actions from the message-passing by introducing new internal channels \code{Enc[i]} and \code{Dec[i]} along with their definitions:
\begin{lstlisting}
add internal family Enc i < n typed: ctxt
 assigned: m : msg <- read In[i];
           k : key <- read Key;
           samp enc((m, k)) then
add internal family Dec i < n typed: msg
 assigned: c : ctxt <- read Enc[i];
           k : key <- read Key;
           return dec((c, k))
\end{lstlisting}
We can now modify the channels \code{Send[i]} and \code{Out[i]} to read from \code{Enc[i]} and \code{Dec[i]} directly:
\begin{lstlisting}
sym from change fam Send with
    e : ctxt <- read Enc[i]; 
    return e
    in currentProtocol(
       subst fam Enc into fam Send ) then
sym from change fam Out with
    okCtxt : unit <- read OkCtxtAdvNet[i];
    d : msg <- read Dec[i]; 
    return d
    in currentProtocol(
       subst fam Dec into fam Out )
\end{lstlisting}
We can now invoke the correctness assumption in each individual session to cancel the effect of encryption followed by decryption. In \cite{ipdl}, the generalization to $n$ sessions was stated as a separate lemma with a nontrivial manual proof (lines 67 -- 103 in CPA.v). In our code, the corresponding proof looks like this:
\begin{lstlisting}
by induction on i with variable x (
  use assumption enc-dec-correctness 
    on chn Dec[x], chn Enc[x] )
\end{lstlisting}
The above code snippet proves by induction on $i < n$ that if 
the channels \code{Dec(i)} with $i<x$ rewrite from the original
formulation  that performs the decryption to the new formulation that simply reads off the message \code{In[i]}, so does the channel \code{Dec[x]}.

On the other hand, our CPA assumption is applied just once across all sessions upon encountering the following protocol snippet:
\begin{lstlisting}
new Key : key in
  ((Key ::= samp gen_key(())) ||
   (family Enc[i] i < n ::= 
      m : msg <- read In[i];
      k : key <- read Key;
      samp enc((m, k))))
\end{lstlisting}
We invoke the approximate CPA assumption as shown below:
\begin{lstlisting}
use approx assumption cpa
\end{lstlisting}
This yields the following protocol snippet:
\begin{lstlisting}
new Key : key in
  ((Key ::= samp gen_key(())) ||
   (family Enc[i] i < n ::=
      m : msg <- read In[i];
      k : key <- read Key;
      samp enc((zeros(()), k))))
\end{lstlisting}
A final step in the proof \emph{folds} the internal channels \code{Enc[i]} and \code{Dec[i]} that we introduced earlier into the rest of the protocol:
\begin{lstlisting}
fold fam Enc into fam LeakCtxtNetAdv then
fold fam Dec into fam Out
\end{lstlisting}
This is justified by the rule \textsc{fold-bind} in Figure~\ref{fig:protocols_equality_strict}, which states that if we only read from channel $c$ once, we can soundly replace $\read{c}$ by the computation assigned to $c$, even if this computation is probabilistic.

\paragraph{Concrete Security Bounds}
Such a proof could be carried out in the IPDL logic alone. However, that proof would only guarantee asymptotic security with respect to the custom symbolic bounds defined in \cite{ipdl}. In this work, we aim for \emph{concrete} security with respect to a low-level Turing Machine semantics: a precise probabilistic bound for the difference in the probability that an adversary can distinguish \code{Real} from \code{Ideal || Sim}.

After encoding the proof in the DSL, our tool computes the following bounds:
\begin{lstlisting}
indistinguishability assumption cpa :
count: 1
context: n * | msg | * 6 + n * | ctxt | * 3 + n * 96 + 12
\end{lstlisting}
Here \code{count} denotes the number of times the CPA assumption was applied, and \code{context} bounds the maximal size of the program context in which it was applied. The expressions \code{|msg|} and \code{|ctxt|} denote the concrete length of bitstrings needed to represent the two types. 

Given these two quantities, our main theorem allows us to derive the following concrete security bound on the distinguishing
advantage for our protocol:
\begin{align*}
& \Big|\mathsf{Pr}\big[\interaction{\Adv}{\mathsf{Real}}{} = 1\big] - \mathsf{Pr}\big[\interaction{\Adv}{\mathsf{Ideal}\ ||\ \mathsf{Sim}}{} = 1\big]\Big| \leq \varepsilon_\mathsf{cpa}.
\end{align*}
Here, $\interaction{\Adv}{P}{}$ denotes the interaction of the adversary $\Adv$ with a protocol $P$. The value $\varepsilon_\mathsf{cpa}$ is the maximal distinguishing advantage for the \code{CPA} assumption against any adversary with computational ``cost'' at most $\mathcal{P}(\text{\code{context}})$. Here $\mathcal{P}$ is a fixed polynomial, and ``cost'' bounds the runtime of the adversary and the number of states in its Turing Machines (among other quantities). In other words, our theorem exactly bounds the security error in our protocol's proof by the error present in the IND-CPA game against the reduction.

Our novel strategy for computing concrete security bounds is to implement an \emph{interpreter} for IPDL programs as a Turing Machine, and compute its cost as the aforementioned polynomial $\mathcal{P}$. The polynomial takes the size of the interpreted program as its argument, and bounds the number of Turing Machine states and transitions that the resulting interpretation will need.

\subsection{The Main Theorem}
Before giving technical details, we now discuss our main result informally. 
Roughly speaking, if $P$ is approximately equal to $Q$, then the advantage that an adversary has in distinguishing $P$ and $Q$ is a reasonable combination of the distinguishing advantages against each indistinguishability assumptions by an adversary whose computational resources are only slightly larger than those of the original adversary.

\begin{theorem*}[Soundness of approximate equality of protocols, informal]\label{thm:main}
There exists a polynomial $\mathcal{P}(x,y,z)$ with the following property. Given:
\begin{itemize}
\item finitely many built-in functions that can be computed with cost at most $C_\sem \in \nat$ and can be approximated by probabilistic Turing Machines with error at most $\eta_\sem \in \rat_{\geq 0}$;
\item indistinguishability assumptions $\vdash P^1 \approx Q^1$, $\ldots$, $\vdash P^n \approx Q^n$;
\item a proof of indistinguishability $\vdash P \approx Q$ with output bounds $\text{\code{count}}_i$ and $\text{\code{context}}_i$ for the $i$-th indistinguishability assumption;
\item an adversary $\Adv$ for $P/Q$ that computes with cost at most $C_\adv \in \nat$;
\item axiom bounds $\varepsilon^1,\ldots,\varepsilon^n \in \rat_{\geq 0}$ with the property that for any adversary $\Adv^i$ for $P^i/Q^i$ such that $\Adv^i$ computes with cost at most $\mathcal{P}(C_\sem,C_\adv,\text{\code{context}}_i)$, we have
\[\Big|\mathsf{Pr}\big[\interaction{\Adv^i}{P^i}{\int{-}} = 1\big] - \mathsf{Pr}\big[\interaction{\Adv^i}{Q^i}{\int{-}} = 1\big]\Big| \leq \varepsilon^i,\]
\end{itemize}
we have
\[ 
\mathsf{Pr}\big[\interaction{\Adv}{P}{\int{-}} = 1\big] - \mathsf{Pr}\big[\interaction{\Adv}{Q}{\int{-}} = 1\big]\Big| \leq 
     \sum_{i = 1}^n \text{\code{count}}_i * \varepsilon^i .
\]
\end{theorem*}

In other words, we can bound the probability in distinguishing $P$ from $Q$ by the sum of the maximal probabilities of violating an indistinguishability axiom. The actual theorem we prove (Thm.~\ref{lem:soundness_approximate}) is slightly more
general, as we allow the distribution symbols to be general distributions that are only \emph{approximated} by Turing Machines, which requires an \emph{error term} to appear in the theorem. 

The probability $\varepsilon^i$ must work for \emph{every} adversary with cost at most $\mathcal{P}(C_\sem, C_\adv, \text{\code{context}}_i)$; thus, the larger the adversary cost, the looser the bound $\varepsilon^i$ must be. Indeed, if $\mathcal{P}(C_\sem, C_\adv, \text{\code{context}}_i) = \text{\code{context}}_i^{500}$, the bound is still polynomial but not particularly meaningful. Crucially, our proof is \emph{constructive}: we are able to
\emph{compute} the polynomial as
\[\mathcal{P}(x, y, z) = y^2 + 8yz + 15z^2 + (|\Sigma_f| + |\Sigma_d| + 1)x + 34y + 47z + O(1).\]
This polynomial, where $|\Sigma_f|$ and $|\Sigma_d|$ are the number of function and distribution symbols in our signature, serves as a precise bound on the reduction overhead incurred when using a cryptographic assumption. We are able to achieve such a concrete bound precisely by using a low-level computational semantics for adversaries as Turing Machines. The exact polynomial, which we include in the appendix, is computed once and for all, and counts the precise number of steps that our TMs take, along with binding the TM's number of states, tapes, and symbols (to encode a protocol on a tape, we use additional symbols besides $0,1$). This is in contrast to almost all prior cryptographic work targeting MPC, which either reasons about runtime informally (without being able to compute the polynomial bound on reduction overhead) or subverts reasoning about runtime via error terms that contain reductions themselves~\cite{ssp}. A notable exception is Barbosa et al.~\cite{advcompl}, which adds a Hoare logic for running time to EasyCrypt, at the cost of being limited to imperative programs (and thus imperative encodings of protocols), and
manual proof effort for each runtime bound.

\subsubsection{Overview of Results}

In this work, we build a new cost-aware proof system on top of the existing exact equational logic of IPDL. The rest of our framework diverges significantly. In particular:

\begin{itemize}
\item We carry out explicit cryptographic reductions instead of using symbolic adversaries. We represent an adversary (Section~\ref{subsec:adversaries}) as a tuple of (essentially arbitrary) probabilistic algorithms for scheduling interactions, updating the adversary's internal state, querying the protocol for output channel values, and assigning new input channel values. When absorbing a program context $Q$ into the adversary, we explicitly extend the adversary's state by the encoding of $Q$ as a sequence of symbols on the Turing Machine tape.

\item We deliver \emph{concrete security bounds} (Section~\ref{subsec:congruence} and~\ref{subsec:approximate}) instead of symbolic ones. In particular, the bound induced by invoking an approximate congruence rule, which allows us to conclude $\Par{P}{Q} \approx \Par{P'}{Q}$ from $P \approx P'$, is the length $\tmnorm{Q}$ of the aforementioned Turing Machine encoding (plus some overhead).

\item We give a natural definition of computational indistinguishability
    (Sec.~\ref{subsec:indistinguishability}) that does not involve reasoning
        about syntactic contexts (as in prior work~\cite{ipdl}). Informally, we define two families of protocols to be indistinguishable if for any polynomial $p(\lambda)$ and negligible function $\eta(\lambda)$, there exists a negligible function $\varepsilon(\lambda)$ such that for any sufficiently large $\lambda$ and any adversary $\Adv$ with cost bounded by $p(\lambda)$, the distinguishing advantage of $\Adv$ is bounded by $\varepsilon(\lambda)$.

\item We carry out an explicit analysis of errors induced by a probabilistic Turing Machine that ends up in a non-accepting state with a negligible probability. This probability makes an appearance in the concrete bounds we derive (Thm.~\ref{lem:soundness_approximate}).
\end{itemize}

\section{Cost-Aware Syntax And Semantics for IPDL}\label{sec:approximate}
\newcommand{\St}{\mathsf{St}}
\newcommand{\Symb}{\mathsf{Symb}}
\newcommand{\Step}{\mathsf{T}}
\renewcommand{\In}{\mathsf{I}}
\renewcommand{\Out}{\mathsf{O}}
\renewcommand{\Dec}{\mathsf{D}}

In this section, we extend the IPDL logic~\cite{ipdl} to handle
\emph{cost-aware} proofs; that is, proofs which guarantee precise concrete security bounds. Our main theorem for concrete security bounds assumes a sound ambient theory for the strict fragment of IPDL. 

While the exact fragment of IPDL is exactly what is desired for cryptographic proofs, the approximate fragment is missing a crucial point of reasoning. In particular, its soundness proof is in terms of \emph{symbolic} bounds, which abstract away the underlying cost semantics of IPDL protocols; in short, it does not reason about runtime. Because of this, the prior approximate logic of IPDL
does not prove the same class of security results generally accepted by the cryptographic community.

For the rest of this section, we assume a fixed signature $\Sigma$ with type constants $\type_1,\ldots,\type_{|\Sigma_\type|}$.

\subsection{Approximate Congruence}\label{subsec:congruence}

Our cost-aware equational theory consists of three layers. Firstly, we have \emph{approximate conguence}, see Figure~\ref{fig:protocols_congruence_approx}, which applies a program context to a single \emph{approximate axiom}, resulting in the judgment $\Delta \vdash \approxcong{P}{Q}{I}{O}{k}{\psi}$. We assume a set of $n$ approximate axioms of the form $\Delta^k \vdash P^k \approx Q^k : I^k \to O^k$ for $1 \leq k \leq n$, where $\Delta^k \vdash P^k : I^k \to O^k$ and $\Delta^k \vdash Q^k : I^k \to O^k$. These axioms capture cryptographic assumptions on computational indistinguishability.

Here, the $\context$ parameter tracks the increase in the adversary's resources incurred by the proof. A typical proof step in the exact
fragment transforms the protocol into a form where an approximate axiom applies. We subsequently carry out an approximate congruence step, where we use the approximate axiom to replace a small protocol fragment nested inside an arbitrary program context by its computationally indistinguishable counterpart.

The program context is formally a part of the adversary, and as such it must be resource-bounded for the indistinguishability assumption to apply. Some nesting patterns do not effect any change on the adversary's resources: for example, a simple renaming of channels (rule \textsc{embed}); the formal addition of an unused channel $i$ to the protocol's inputs $I$ (rule \textsc{input-unused}), in which case any value assigned by the adversary to channel $i$ will leave the protocol unchanged; or the introduction of an internal channel $o : \tau$ (rule \textsc{cong-new}), in which case the adversary will never query $o$ because internal channels are only visible in the scope of their declaration.

On the other hand, composing two approximately equal protocols $P \approx P'$ with another protocol $Q$ requires the adversary to simulate the interaction of the program context $Q$ with $P$ versus $P'$. In other words, the adversary \emph{absorbs} $Q$ and the protocol becomes part of the new adversary's code. In particular, the number of symbols needed for encoding the adversary's code on a
Turing Machine tape increases, and the parameter $\psi$ approximates this increase. As rule \textsc{cong-comp} shows, composition with protocol $Q$ incurs $\tmnorm{Q} + 3$ additional symbols: $\tmnorm{Q}$ symbols for encoding $Q$; a parallel composition symbol to combine the original code with the code for $Q$; and two parenthesis symbols \textsf{``$\mathsf{(}$''}, \textsf{``$\mathsf{)}$''} for enclosing the composition. We emphasize that the exact numbers here are not crucial; what matters is that we eventually deliver a (reasonable) polynomial in $\lambda$.

We show how to compute the Turing Machine bound of an IPDL construct in Section~\ref{sec:tm_bound}. This bound, and consequently the $\context$ parameter $\psi$, is not a natural number but a function $\psi(t_1,\ldots,t_{|\Sigma_\type|}) : \nat^{|\Sigma_\type|} \to \nat$ that is \emph{monotonically increasing in each argument}. When encoding a protocol $Q$ as a sequence of symbols on a Turing Machine tape, we invariably encounter variables $x$ of type $\tau$. At this point, we do not know how many bits we will need to encode values of type $\tau$, because the type constants $\type \in \Sigma$ are yet uninterpreted. Instead, we leave the size of each type constant as a variable to the function $\psi$, which will later be instantiated by the appropriate natural number according to $\int{-}$. 

\subsection{Approximate and Asymptotic Equality}\label{subsec:approximate}
In the \emph{approximate equality} of protocols, see Figure~\ref{fig:protocols_equality_approx}, we chain together a sequence of strict equalities and approximate congruence transformations
to obtain the judgment $\Delta \vdash \approxeq{P}{Q}{I}{O}{\xi}{\psi}$.
The parameter $\xi$ counts the number of axiom invocations for each of the $n$ axioms. One application of the $k$-th approximate axiom incurs a count that maps $k$ to $1$ and all other axioms to $0$. The use of transitivity requires us to add up the respective values of $\xi$ per each axiom (rule \textsc{trans}). Even though each individual axiom invocation introduces a negligible error, summing up exponentially many negligible errors might not be negligible, which is why we need to keep track of the number of times each axiom is applied. The parameter $\psi$ tracks the maximum size of a program context in which each axiom is applied.

\begin{figure*}
\begin{mathpar}
\fbox{$\Delta \vdash \approxcong{P}{Q}{I}{O}{k}{\psi}$}\\
\inferrule*[right=axiom]{ }{\Delta^k \vdash \approxcong{P^k}{Q^k}{I^k}{O^k}{k}{0}}\and
\inferrule*[right=input-unused]{c \notin I \cup O \\ \Delta \vdash \approxcong{P}{Q}{I}{O}{k}{\psi}}{\Delta \vdash \approxcong{P}{Q}{I \cup \{c\}}{O}{k}{\psi}}\and
\inferrule*[right=embed]{\phi : \Delta_1 \to \Delta_2 \\ \Delta_2 \vdash \approxcong{P}{Q}{I}{O}{k}{\psi}}{\Delta_1 \vdash \approxcong{\phi^\star(P)}{\phi^\star(Q)}{\phi^\star(I)}{\phi^\star(O)}{k}{\psi}}\and
\inferrule*[right=cong-comp]{\Delta \vdash \approxcong{P}{P'}{I \cup O_2}{O_1}{k}{\psi} \\ \Delta \vdash Q : I \cup O_1 \to O_2}{\Delta \vdash \approxcong{\Par{P}{Q}}{\Par{P'}{Q}}{I}{O_1 \cup O_2}{k}{(\psi + \tmnorm{Q} + 3)}}\and
\inferrule*[right=cong-new]{\Delta, o : \tau \vdash \approxcong{P}{P'}{I}{O \cup \{o\}}{k}{\psi}}{\Delta \vdash \approxcong{\big(\new{o}{\tau}{P}\big)}{\big(\new{o}{\tau}{P'}\big)}{I}{O}{k}{\psi}}
\end{mathpar}
\caption{Approximate congruence of IPDL protocols.}
\label{fig:protocols_congruence_approx}
\end{figure*}

\begin{figure*}
\begin{mathpar}
\fbox{$\Delta \vdash \approxeq{P}{Q}{I}{O}{\xi}{\psi}$}\\
\inferrule*[right=strict]{\Delta \vdash P = Q : I \to O}{\Delta \vdash \approxeq{P}{Q}{I}{O}{(i \mapsto 0)}{(i \mapsto 0)}}\and
\inferrule*[right=approx-cong]{\Delta \vdash \approxcong{P}{Q}{I}{O}{k}{\psi}}{\Delta \vdash \approxeq{P}{Q}{I}{O}{\big(k \mapsto 1, i \neq k \mapsto 0\big)}{\big(k \mapsto \psi, i \neq k \mapsto 0\big)}}\and
\inferrule*[right=sym]{\Delta \vdash \approxeq{P_1}{P_2}{I}{O}{\xi}{\psi}}{\Delta \vdash \approxeq{P_2}{P_1}{I}{O}{\xi}{\psi}}\and
\inferrule*[right=trans]{\Delta \vdash \approxeq{P_1}{P_2}{I}{O}{\xi_1}{\psi_1} \\ \Delta \vdash \approxeq{P_2}{P_3}{I}{O}{\xi_2}{\psi_2}}{\Delta \vdash \approxeq{P_1}{P_3}{I}{O}{\big(i \mapsto \xi_1(i) + \xi_2(i)\big)}{\big(i \mapsto \max(\psi_1(i), \psi_2(i))\big)}}
\end{mathpar}
\caption{Approximate equality for IPDL protocols.}
\label{fig:protocols_equality_approx}
\end{figure*}

Finally, analogously to \cite{ipdl}, we define the \emph{asymptotic equality} of two protocol families, see Figure~\ref{fig:protocols_equality_asympto}, as functions of the security parameter $\lambda \in \nat$. Informally speaking, if two protocol families are asymptotically equal, then any \emph{resource-bounded} adversary cannot distinguish them with greater than negligible error. Formally, we assume a finite set of \emph{approximate axiom families} of the form $\big\{\Delta_\lambda \vdash P_\lambda \approx Q_\lambda : I_\lambda \to O_\lambda\big\}_{\lambda \in \nat}$, where $\{P_\lambda\}, \{Q_\lambda\}$ are two protocol families with pointwise-identical typing judgments. If the axiom families comprising our asymptotic theory are clear from the context, we will omit them from the asymptotic equality judgment.

For any fixed $\lambda$, we obtain an approximate theory by selecting from each axiom family the particular axiom corresponding to $\lambda$. Similarly, from each of the two protocol families we select the protocol corresponding to $\lambda$, which gives us two concrete protocols to equate approximately. We recall that an approximate equality judgment is tagged by a pair of parameters $\xi$ and $\psi$, and for each axiom $1 \leq i \leq n$ we have $\xi^i \in \nat$ and $\psi^i : \nat^{|\Sigma_\type|} \to \nat$, where $|\Sigma_\type|$ is the number of type constants declared in our ambient signature $\Sigma$. Fixing the axiom $i$ and letting the security parameter $\lambda$ vary thus gives us two functions $\nat \to \nat$ and $\nat^{|\Sigma_\type|+1} \to \nat$, and we require that these be bounded by polynomials in the appropriate number of variables. Unlike \cite{ipdl}, we do not impose a bound on the channel context $\Delta$, which can be particularly burdensome to check in a formal tool. Instead, our definition of an adversary ensures that it only interacts with the protocol via a bounded number of channels.

\begin{figure*}
\begin{mathpar}
\fbox{$\big\{\Delta^1_\lambda \vdash P^1_\lambda \approx Q^1_\lambda : I^1_\lambda \to O^1_\lambda\big\}_{\lambda}, \ldots, \big\{\Delta^n_\lambda \vdash P^n_\lambda \approx Q^n_\lambda : I^n_\lambda \to O^n_\lambda\big\}_{\lambda} \; \mathlarger{\mathlarger{\mathlarger{\vdash}}} \; \big\{\Delta_\lambda \vdash P_\lambda \approx Q_\lambda : I_\lambda \to O_\lambda\big\}_{\lambda}$}\\
\inferrule{\forall \lambda, \Delta^i_\lambda \vdash P^i_\lambda \approx Q^i_\lambda : I^i_\lambda \to O^i_\lambda, i = 1, \ldots, n\; \mathlarger{\mathlarger{\vdash}} \; \Delta_\lambda \vdash \approxeq{P_\lambda}{Q_\lambda}{I_\lambda}{O_\lambda}{\xi_\lambda}{\psi_\lambda} \\ \forall i, \xi_{(\cdot)}^i = \mathsf{O}(\poly(\lambda)) \\ \forall i, \psi_{(\cdot)}^i = \mathsf{O}\big(\poly(\lambda,t_1,\ldots,t_{|\Sigma_\type|})\big)}{\big\{\Delta^1_\lambda \vdash P^1_\lambda \approx Q^1_\lambda : I^1_\lambda \to O^1_\lambda\big\}_{\lambda}, \ldots, \big\{\Delta^n_\lambda \vdash P^n_\lambda \approx Q^n_\lambda : I^n_\lambda \to O^n_\lambda\big\}_{\lambda} \; \mathlarger{\mathlarger{\vdash}} \; \big\{\Delta_\lambda \vdash P_\lambda \approx Q_\lambda : I_\lambda \to O_\lambda\big\}_{\lambda}}
\end{mathpar}
\caption{Asymptotic equality for IPDL protocol families.}
\label{fig:protocols_equality_asympto}
\end{figure*}

\subsection{Adversaries for IPDL Protocols}\label{subsec:adversaries}

To seamlessly account for the possible renaming of channel names, an adversary for protocols of type $\Delta \vdash I \to O$ is allowed to operate in a larger context $\Delta'$ that subsumes the original context $\Delta$ via an \emph{embedding} $\phi : \Delta'\to \Delta$. Here $\phi$ is an injective, type-preserving mapping that specifies how to rename channels in $\Delta$ to fit in the larger context $\Delta'$. In this larger context, we specify the adversarial input channels $I'$ that the adversary will query for a value, and the adversarial output channels $O'$ that the adversary will assign values to. The adversarial inputs $I'$ will be a subset of the protocol outputs $O$, appropriately translated along $\phi$. Dually, the protocol inputs $I$, appropriately translated along $\phi$, will be a subset of the adversarial outputs $O'$. In the interaction between the adversary and the protocol, every query for a value of a channel $o \in I'$ will extract the value of the channel $o \in \phi^\star(O)$ as computed by the the protocol, and pass it on to the adversary. Conversely, an input on channel $i \in \phi^\star(I)$ to the protocol occurs after the adversary computes the value of the channel $i \in O'$.

Since our adversaries will be resource-bounded, we need to bind the number of interactions or \emph{rounds} between the adversary and the protocol. In each round, the adversary examines its internal state to determine the type of interaction to perform next, and steps to a new state. This transition function is a partial probabilistic function of type $\St \rightharpoonup \big(\{\bot\} \cup I' \cup O'\big) \times \St$. That is, for any internal state $s$ the adversary probabilistically decides among: \emph{1)} no interaction, coupled with stepping to a new state $s'$; \emph{2)} querying a channel $o \in I'$, coupled with stepping to a new state $s'$; \emph{3)} an assignment to a channel $i \in O'$, coupled with stepping to a new state $s'$; or \emph{4)} halting, in which case the game between the adversary and the protocol ends without a decision Boolean. We use this last option to capture probabilistic computations that only succeed up to a negligible error.

If the adversary queries channel $o \in I'$ and receives a value $v$ as a response to the query, it updates its internal state according to an input assignment function of type $\int{\tau} \times \St \to \St$, where $\tau$ is the type of the channel $o$ in $\Delta'$. That is, for any value $v \in \int{\tau}$ and any state $s$, the adversary steps to a new state $s'$ that records the value $v$ as a result of the query. If the adversary chooses a value assignment to a channel $i \in O'$, the value $v$ -- if any -- is determined by an output valuation function of type $\St \to \int{\tau} \cup \{\bot\}$, where $\tau$ is the type of the channel $i$ in $\Delta'$. After completing the designated number of rounds, the adversary converts its internal state to a final decision Boolean according to a function of type $\St \to \{0,1\}$.

To bind the complexity of the aforementioned operations, we implement them as Turing Machines\footnote{When using Turing Machines to compute (probabilistic) functions, we only consider (probabilistic) Turing Machines that have a finite runtime $N \in \nat$; \emph{i.e.}, for every input in the domain, after $N$ (probabilistic) transitions the TM ends up in a configuration where no further transitions are possible. That is, the TM has either reached an accepting state or it has halted after reading a symbol for which no transition is possible in the current state.}. For convenience, we allow TMs with multiple tapes. As is standard, in the initial configuration all tapes except the first are fully blank. The internal state of the adversary is typically encoded as a bitstring, containing \emph{e.g.}, register values together with the sequence of instructions to be executed, if we wish to view the adversary as an essentially arbitrary probabilistic program. For our purposes, it is convenient to allow additional symbols besides $0, 1$ on our TM tapes: when justifying the \textsc{comp-cong} rule of our theory, we suitably compose the adversary with the common context $Q$. The protocol $Q$ thus becomes integrated into the new adversary's code. Instead of encoding protocols as bitstrings, we will suitably enrich our baseline set of symbols so that we can faithfully capture IPDL code.

\begin{definition}[Adversary]\label{def:disting}
Fix an interpretation $\int{-}$ for $\Sigma$. An \emph{adversary} for protocols $\Delta \vdash I \to O$ is a tuple $\big(\Delta', I', O', \phi, \#_\round,\#_\tape, \Symb, \St, s_\star, \Step, \big\{\In_o\big\}_{o \, \in \, I'}, \big\{\Out_i\big\}_{i \, \in \, O'}, \Dec\big)$, where

\begin{itemize}
\item $\Delta'$ is a channel context;

\item $I' \subseteq \Delta'$ is a set of channels that the adversary can query for a value;

\item $O' \subseteq \Delta'$ is a set of channels to which the adversary can assign a value;

\item $\phi : \Delta' \to \Delta$ is an embedding of $\Delta$ into $\Delta'$;

\item $\#_\round \geq 1$ is the number of rounds the adversary will perform;

\item $\#_\tape \geq 1$ is the number of TM tapes at our disposal;

\item $\Symb$ is a finite set of additional symbols that will be used to encode the adversary's internal state;

\item $\St \subseteq \big(\{0,1\} \, \bigsqcup \, \Symb\big)^l$ is a set of strings of a fixed length $l \geq 1$ consisting of symbols drawn from the disjoint union of the sets $\{0,1\}$ and $\Symb$;

\item $s_\star \in \St$ is the initial state;

\item $\Step$ is a probabilistic TM that computes a partial function $\St \rightharpoonup \big(\{\bot\} \cup I' \cup O'\big) \times \St$, with $\#_\tape$-many tapes using symbols from the set $\{0,1\} \, \bigsqcup \ \{\bot\} \, \bigsqcup \ (I' \cup O') \, \bigsqcup \, \Symb$,

\item $\In_o$ where $o : \tau \in \Delta$ is a TM that computes a function $\St \times \int{\tau} \to \St$, with $\#_\tape$-many tapes using symbols from the set $\{0,1\} \, \bigsqcup \, \Symb$,

\item $\Out_{i}$ where $i : \tau \in \Delta$ is a TM that computes a function $\St \to \int{\tau} \cup \{\bot\}$, with $\#_\tape$-many tapes using symbols from the set $\{0,1\} \, \bigsqcup \ \{\bot\} \, \bigsqcup \, \Symb$,

\item $\Dec$ is a TM that computes a function $\St \to \{0,1\}$, with $\#_\tape$-many tapes using symbols from the set $\{0,1\} \, \bigsqcup \, \Symb$.
\end{itemize}
We furthermore require that $I' \subseteq \phi^\star(O)$, $\phi^\star(I) \subseteq O'$, and $\phi^\star(O) \cap O' = \emptyset$.
\end{definition}

\noindent The probabilistic TM $\Step$ that computes the transition function can terminate in a non-accepting state with probability $> 0$. We will be interested in families of adversaries where this \emph{error} as a function of the security parameter is negligible.

\begin{definition}[Adversarial error]
We say that an adversary $\Adv$ \emph{has error up to} $\varepsilon \in \rat_{\geq 0}$, written $\err(\Adv) \leq \varepsilon$, if for any state $s \in \St$ the transition function $\Step(s)$ is undefined with probability $\leq \varepsilon$. In other words, when $\Step$ is run with the initial tape contents $s$, it halts in a non-accepting state with probability $\leq \varepsilon$.
\end{definition}

\noindent To ensure that the adversary does not have access to computationally expensive functions such as the discrete logarithm, we need to impose a bound on its computational resources. We will be interested in families of adversaries where the bound as the function of the security parameter is polynomial.

\begin{definition}[Resource-bounded adversaries]
We say that an adversary $\Adv$ has \emph{cost at most} $K \in \nat$, written $|\Adv| \leq K$, if:
\begin{itemize}
\item $\#_\round$, $\#_\tape \leq K$;
\item $|I'| \leq K$ and for each $o \in I'$ with $o : \tau \in \Delta'$, we have $|\tau| \leq K$,
\item $|O'| \leq K$ and for each $i \in O'$ with $i : \tau \in \Delta'$, we have $|\tau| \leq K$,
\item $|\Symb| \leq K$;
\item the length $k$ of a state $s \in \St$ is $\leq K$;
\item the number of states of each\footnote{Instead of having a separate Turing Machine $\In_o$ for each channel $o \in I'$ we could have required a single Turing Machine that performs the computation across all channels in $I'$, and analogously for $\Out_i$. However, this is unnecessary as the number of channels in both $I'$ and $O'$ is $\mathsf{O(poly(\lambda))}$, and the current formulation is more convenient for our purposes.} TM $\Step, \In_o, \Out_i, \Dec$ is $\leq K$;
\item the runtime of each TM $\Step, \In_o, \Out_i, \Dec$ is $\leq K$.
\end{itemize}
\end{definition}

\begin{definition}[Interaction]
Fix an interpretation $\int{-}$ for $\Sigma$. Let $\Adv$ be an adversary for protocols $\Delta \vdash I \to O$ and let $\Delta \vdash P : I \to O$. We define $\interaction{\Adv}{P}{\int{-}}$ to be the probability sub-distribution on Booleans induced by the algorithm in Figure~\ref{fig:interaction}.
\end{definition}

\begin{figure}[ht]
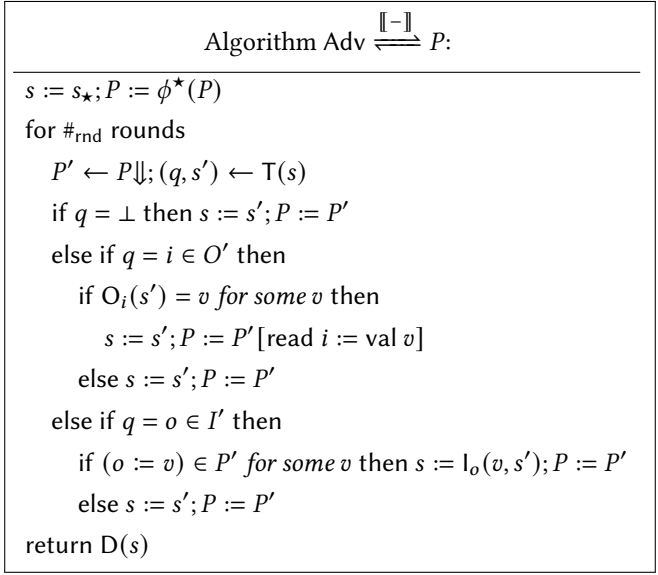

\fbox{% 
\begin{minipage}{0.6 \textwidth}
\begin{center}
Algorithm $\interaction{\Adv}{P}{\int{-}}$:
\end{center}
\hrule
\begin{align*}
& s := s_\star ; P := \phi^\star(P) \\
& \mathsf{for \ } \#_\round \ \mathsf{rounds} \\
& \ \ \ \ P' \leftarrow \eval{P} ; (q, s') \leftarrow \Step(s) \\
& \ \ \ \ \mathsf{if} \ q = \bot \ \mathsf{then} \ s := s' ; P := P' \\
& \ \ \ \ \mathsf{else \ if} \ q = i \in O' \ \mathsf{then} \\
& \ \ \ \ \ \ \ \ \mathsf{if} \ \Out_i(s') = v \ \textit{for some $v$} \ \mathsf{then} \\
& \ \ \ \ \ \ \ \ \ \ \ \ s := s' ; P := P'[\read{i} := \val{v}] \\
& \ \ \ \ \ \ \ \ \mathsf{else} \ s := s' ; P := P'\\
& \ \ \ \ \mathsf{else \ if} \ q = o \in I' \ \mathsf{then} \\
& \ \ \ \ \ \ \ \ \mathsf{if} \ (\assign{o}{v}) \in P' \ \textit{for some $v$} \ \mathsf{then} \ s := \In_o(v,s') ; P := P' \\
& \ \ \ \ \ \ \ \ \mathsf{else} \ s := s' ; P := P' \\ 
& \mathsf{return} \ \Dec(s)
\end{align*}%
\end{minipage}
}
\caption{Interaction of an an adversary $\Adv$ with an IPDL protocol $P$.}
\label{fig:interaction}
\end{figure}

In Figure~\ref{fig:interaction}, the adversary interacts with the protocol through the specified number of rounds. The algorithm maintains a state variable $s$ along with a protocol variable $P$, which we respectively initialize to the initial state $s_\star$ and the original protocol $P$, appropriately embedded in $\Delta'$. In each round, the protocol $P$ probabilistically evolves to a new protocol $P'$. Independently, the adversary probabilistically computes the type of interaction $q \in \{\bot\} \cup I' \cup O'$ together with a new state $s'$ according to $\Step(s)$. If $q = \bot$, in which case no interaction takes place, the state and the protocol are updated to $s'$ and $P'$. If $q = i$ for some $i \in O'$, we compute $\Out_i(s')$ to see if in the adversary's current state $s'$ the channel $i$ carries a value $v$. If this is the case, we update the state to $s'$ while computing a new protocol $P'[\read{i} := \val{v}]$. Otherwise we update the state and the protocol to $s'$ and $P'$. Finally, if $q = o$ for some $o \in I'$, we examine the protocol $P'$ to see if the output channel $o$ carries a value $v$. If this is the case, we compute a new adversary state $\In_o(v,s')$, while updating the protocol to $P'$. Otherwise we update the state and the protocol to $s'$ and $P'$. After completing the prescribed number of rounds, we obtain a decision Boolean\footnote{Strictly speaking, the interaction $\interaction{\Adv}{P}{\int{-}}$ is only a \emph{sub}-distribution on Booleans, since $\Step(s)$ may halt without a result. As the probability of this happening will be negligible, we gloss over this technical point here.} $\Dec(s)$ based on the adversary's current state.

\subsection{Computational Indistinguishability}\label{subsec:indistinguishability}

A family of interpretations is PPT (probabilistic polynomial-time) if it assigns polynomial lengths to type symbols $\type$, and PPT-computable functions to function symbols $\func$ and distribution symbols $\dist$. A small caveat is that a random distribution on a subset $S \subseteq \{0,1\}^n$ of bitstrings is in general computable by a probabilistic Turing Machine only up to a small error $\varepsilon$, which is the probability that the TM does not end up in an accepting state. In effect, the TM computes a distribution $\mu$ on $S \cup \{\bot\}$ with $\mu(\bot) = \varepsilon$. To relate $\mu$ to our original distribution on $S$, we introduce the following:

\begin{definition}[Approximating distributions]
Let $S \subseteq \{0,1\}^n$ be a subset of bitstrings of a fixed length. We say that a distribution $\mu_1$ on $S \cup \{\bot\}$ \emph{approximates} a distribution $\mu_2$ on $S$ with error $0 \leq \varepsilon \leq 1$ if there are distributions $\eta_1, \eta_2$ on $S$ such that $\mu_1 = (1 - \varepsilon) \eta_1 + \varepsilon 1[\bot]$ and $\mu_2 = (1 - \varepsilon) \eta_1 + \varepsilon \eta_2$.
\end{definition}

A function $\varepsilon : \nat \to \rat_{\geq 0}$ is negligible if it is eventually smaller than the inverse of any polynomial: \emph{for any $n \in \nat$, there exists $N \in \nat$ such that for all $\lambda \geq N$ we have $\varepsilon(\lambda) \leq \frac{1}{\lambda^n}$}.

\begin{definition}[PPT family of interpretations]\label{def:ppt-interpretation}
We say that a family $\big\{\int{-}_\lambda\big\}_{\lambda \in \nat}$ of interpretations for $\Sigma$ is \emph{PPT} if there is a polynomial $p(\lambda)$, a negligible function $\eta(\lambda)$, and a natural number $N \in \nat$ such that the following holds:
\begin{itemize}
\item For all type symbols $\type$, $|\type|_\lambda \leq p(\lambda)$ if $\lambda \geq N$.

\item For all function symbols $\func : \sigma \to \tau$, the function $\int{\func}_\lambda$ from bitstrings $\int{\sigma}_\lambda$ to bitstrings $\int{\tau}_\lambda$ is computable by a deterministic Turing Machine $\TM_\lambda$ with symbols $\mathsf{0}, \mathsf{1}$. Both the number of states and the runtime of $\TM_\lambda$ are $\leq p(\lambda)$ if $\lambda \geq N$.

\item For all distribution symbols $\dist : \sigma \to \tau$, the function $\int{\dist}_\lambda$ from bitstrings $\int{\sigma}_\lambda$ to distributions on bitstrings $\int{\tau}_\lambda$ is computable up to an error $\eta(\lambda)$ by a probabilistic Turing Machine $\TM_\lambda$ with symbols $\mathsf{0}, \mathsf{1}$. Specifically, for every $v \in \int{\sigma}_\lambda$, the distribution $\TM_\lambda(v)$ on $\int{\tau}_\lambda \cup \{\bot\}$ approximates $\int{\dist}_\lambda(v)$ with error $\leq \eta(\lambda)$. Both the number of states and the runtime of $\TM_\lambda$ are $\leq p(\lambda)$ if $\lambda \geq N$.
\end{itemize}
\end{definition}

We are now ready to give the definition of computational indistinguishability. We will be only interested in indistinguishability with respect to a PPT family of interpretations.

\begin{definition}[Computational Indistinguishability]\label{def:comp_indist}
Consider a family $\big\{\int{-}_\lambda\big\}_{\lambda \in \nat}$ of interpretations for $\Sigma$. Let $\big\{\Delta_\lambda \vdash P_\lambda : I_\lambda \to O_\lambda\big\}_{\lambda \in \nat}$ and $\big\{\Delta_\lambda \vdash Q_\lambda : I_\lambda \to O_\lambda\big\}_{\lambda \in \nat}$ be two protocol families with identical typing judgments. We say that $\{P_\lambda\}$ and $\{Q_\lambda\}$ are \emph{indistinguishable} under $\big\{\int{-}_\lambda\big\}$, written
\[\big\{\int{-}_\lambda\big\}_{\lambda \in \nat} \; \mathlarger{\mathlarger{\vDash}} \; \big\{\Delta_\lambda \vdash P_\lambda \approx Q_\lambda : I_\lambda \to O_\lambda\big\}_{\lambda \in \nat}\]
if for any polynomial $p(\lambda)$ and negligible function $\eta(\lambda)$, there exists a negligible function $\varepsilon(\lambda)$ and an $N \in \nat$ such that for any $\lambda \geq N$ and any adversary $\Adv$ for protocols $\Delta_\lambda \vdash I_\lambda \to O_\lambda$ with respect to the interpretation $\int{-}_\lambda$ such that $|\Adv| \leq p(\lambda)$ and $\err(\Adv) \leq \eta(\lambda)$, we have\footnote{Instead of comparing probabilities for $1$, we could have likewise used $0$: the probability $\mathsf{Pr}[b = 0]$ that the decision Boolean $b$ is $0$ is only negligibly different from $1 - \mathsf{Pr}[b = 1]$, since the probability that the game ends without a decision Boolean is negligible. This follows from the fact that the error is negligible and the number of rounds is $\mathsf{O(poly(\lambda))}$.}
\[\Big|\mathsf{Pr}\big[\interaction{\Adv}{P_\lambda}{\int{-}_\lambda} = 1\big] - \mathsf{Pr}\big[\interaction{\Adv}{Q_\lambda}{\int{-}_\lambda} = 1\big]\Big| \leq \varepsilon(\lambda).\]
\end{definition}

If the family of interpretations is clear from the context, we may omit it from the computational indistinguishability judgment.

\subsection{Concrete and Asymptotic Security}

We can now present our main result in full:

\begin{theorem}[Soundness of approximate equality of protocols]\label{lem:soundness_approximate}
There exists a polynomial $\mathcal{P}(x,y,z)$ such that for any
\begin{itemize}
\item interpretation $\int{-}$ for $\Sigma$ for which there are $C_\sem \in \nat$ and $\eta_\sem \in \rat_{\geq 0}$ such that
\begin{itemize}
\item for all type symbols $\type$, $|\type| \leq C_\sem$;

\item for all function symbols $\func$, $\int{\func}$ is computable by a TM with symbols $\mathsf{0}, \mathsf{1}$ such that the number of states and the runtime are $\leq C_\sem$; and

\item for all distribution symbols $\dist$, $\int{\dist}_\lambda$ is computable up to an error $\eta_\sem$ by a probabilistic TM with symbols $\mathsf{0}, \mathsf{1}$ such that the number of states and the runtime are $\leq C_\sem$;
\end{itemize}

\item approximate axioms $\Delta^1 \vdash P^1 \approx Q^1 : I^1 \to O^1$, $\ldots$, $\Delta^n \vdash P^n \approx Q^n : I^n \to O^n$;

\item derivation $\Delta \vdash \approxeq{P}{Q}{I}{O}{\xi}{\psi}$;

\item adversary $\Adv$ for protocols of type $\Delta \vdash I \to O$ such that $|\Adv| \leq C_\adv$ and $\err(\Adv) \leq \eta_\adv$ for some $C_\adv \in \nat$ and $\eta_\adv \in \rat_{\geq 0}$,

\item context bounds $C^1_\context,\ldots,C^n_\context \in \nat$ such that $\psi^i(|\type_1|,\ldots,|\type_{|\Sigma_\mathsf{t}|}|) \leq C^i_\context$; and

\item axiom bounds $\varepsilon^1,\ldots,\varepsilon^n \in \rat_{\geq 0}$ with the property that for any adversary $\Adv^i$ for protocols $\Delta^i \vdash I^i \to O^i$ such that \[|\Adv^i| \leq \mathcal{P}(C_\sem,C_\adv,C^i_\context)\] and $\err(\Adv^i) \leq \max(\eta_\sem,\eta_\adv)$, we have
\[\Big|\mathsf{Pr}\big[\interaction{\Adv^i}{P^i}{\int{-}} = 1\big] - \mathsf{Pr}\big[\interaction{\Adv^i}{Q^i}{\int{-}} = 1\big]\Big| \leq \varepsilon^i,\]
\end{itemize}
we have
\[\Big|\mathsf{Pr}\big[\interaction{\Adv}{P}{\int{-}} = 1\big] - \mathsf{Pr}\big[\interaction{\Adv}{Q}{\int{-}} = 1\big]\Big| \leq \sum_{i = 1}^n \xi^i * \big(\varepsilon^i + 2 * C^i_\context * \eta_\sem\big).\]
\end{theorem}

We briefly explain where the error term $2 * C^i_\context * \eta_\sem$ comes from. The original error $\eta_\sem$ is multiplied by $C^i_\context$ because $C^i_\context$ provides an upper bound on how many times we perform probabilistic samplings when executing the absorbed protocol, and hence serves as a proxy for how much the total error accumulates. Finally, we multiply the overall error term by two because computing the distinguishing advantage of an adversary $A$ for protocols $\Par{P_1}{Q}$ versus $\Par{P_2}{Q}$, we accumulate errors twice: once when comparing $\interaction{A}{(\Par{P_1}{Q})}{}$ against $\interaction{B}{P_1}{}$ (where $B$ is the reduced adversary obtained from $A$ by absorbing the program context $Q$), and then again when comparing $\interaction{B}{P_2}{}$ against $\interaction{A}{(\Par{P_2}{Q})}{}$. Our proof of Theorem~\ref{lem:soundness_approximate} relies on two intermediate results. The first one says that strict equality of protocols implies perfect indistinguishability against any adversary (not just a resource-bounded one); for proof see Section~\ref{sec:proof-lemma1}.

\begin{lemma}[Perfect indistinguishability]\label{lem:soundness_approximate_perfect}
For any interpretation $\int{-}$ for $\Sigma$, derivation $\Delta \vdash P = Q : I \to O$, and adversary $\Adv$ for protocols $\Delta \vdash I \to O$, we have
\[\Big|\mathsf{Pr}\big[\interaction{\Adv}{P}{\int{-}} = 1\big] - \mathsf{Pr}\big[\interaction{\Adv}{Q}{\int{-}} = 1\big]\Big| = 0.\]
\end{lemma}

The next result is the \emph{absorption lemma}, the proof of which we sketch in Section~\ref{sec:absorption}, which allows us to absorb a protocol into an adversary at the cost of correspondingly increasing its cost and error. This is precisely where the polynomial $\mathcal{P}(x,y,z)$ in Theorem~\ref{lem:soundness_approximate} comes from:

\begin{lemma}[Absorption]\label{lem:absorption}
There exists a polynomial $\mathcal{P}(x,y,z) \geq y$ such that for any
\begin{itemize}
\item interpretation $\int{-}$ for $\Sigma$ for which there are $C_\sem \in \nat$, $\eta_\sem \in \rat_{\geq 0}$ such that
\begin{itemize}
\item for all type symbols $\type$, $|\type| \leq C_\sem$,

\item for all function symbols $\func$, $\int{\func}$ is computable by a TM with symbols $\mathsf{0}, \mathsf{1}$ such that the number of states and the runtime are $\leq C_\sem$, and

\item for all distribution symbols $\dist$, $\int{\dist}$ is computable up to error $\eta_\sem$ by a probabilistic TM with symbols $\mathsf{0}, \mathsf{1}$ such that the number of states and the runtime are $\leq C_\sem$,
\end{itemize}

\item adversary $\Adv$ for protocols of type $\Delta \vdash I \to O_1 \cup O_2$ such that $|\Adv| \leq C_\adv$ and $\err(\Adv) \leq \eta_\adv$ for some $C_\adv \in \nat$ and $\eta_\adv \in \rat_{\geq 0}$,

\item protocol $\Delta \vdash Q : I \cup O_1 \to O_2$,
\end{itemize}
we have an adversary $\Adv_\mathcal{R}$ for protocols $\Delta \vdash I \cup O_2 \to O_1$ with
\[|\Adv_\mathcal{R}| \leq \mathcal{P}\big(C_\sem,C_\adv,\tmnorm{Q}(|\type_1|,\ldots,|\type_{|\Sigma_\mathsf{t}|}|)\big)\]
and $\err(\Adv_\mathcal{R}) \leq \max(\eta_\sem,\eta_\adv)$ such that for any protocol $\Delta \vdash P : I \cup O_2 \to O_1$ we have
\[\Big|\mathsf{Pr}\big[\interaction{\Adv}{\Par{P}{Q}}{\int{-}} = 1\big] - \mathsf{Pr}\big[\interaction{\Adv_\mathcal{R}}{P}{\int{-}} = 1\big]\Big| \leq \tmnorm{Q}(|\type_1|,\ldots,|\type_{|\Sigma_\mathsf{t}|}|) * \eta_\sem.\]
\end{lemma}

The high-level idea behind the proof of Theorem~\ref{lem:soundness_approximate} is to restructure the derivations of approximate equality so that all invocations of the rule \textsc{embed} are carried out first, followed by applications of the rule \textsc{input-unused}, which are in turn followed by invocations of the rule \textsc{cong-comp}, and finally by applications of the rule \textsc{cong-new}. The new layered form of our approximate judgments is shown in 
Figures~\ref{fig:protocols_layered_1} and~\ref{fig:protocols_layered_2}.

Crucially, we collapse a sequence of applications of the \textsc{cong-comp} rule with common contexts $Q_1, \ldots, Q_n$ into a single application with the combined common context $Q_1  \; || \; \ldots \; || \; Q_n$. This is necessary because every time we absorb a protocol into the adversary, we increase the adversary's resources: \emph{e.g.}, the number of rounds that $\Adv_\mathcal{R}$ takes is more than double the original number of rounds. Thus, if we carried out this process $\lambda$-many times, we would see an exponential increase in the adversary's resources.

Our second result (Theorem~\ref{thm:soundness_asympto}; for proof see Section~\ref{sec:soundness_asympto}) concerns asymptotic security and serves as a sanity check for the concrete bounds we derived. Our asymptotic theory is said to be sound if each of its axioms is sound:

\begin{definition}
We say that an approximate axiom family $\big\{\Delta_\lambda \vdash P_\lambda \approx Q_\lambda : I_\lambda \to O_\lambda\big\}_{\lambda \in \nat}$ is \emph{sound} with respect to a family of interpretations $\big\{\int{-}_\lambda\big\}_{\lambda \in \nat}$ if $\big\{\int{-}_\lambda\big\}_{\lambda \in \nat} \; \mathlarger{\mathlarger{\vDash}} \; \big\{\Delta_\lambda \vdash P_\lambda \approx Q_\lambda : I_\lambda \to O_\lambda\big\}_{\lambda \in \nat}$.
\end{definition}

\begin{theorem}[Soundness of asymptotic equality of protocols]\label{thm:soundness_asympto}
For any
\begin{itemize}
\item protocol families $\big\{\Delta_\lambda \vdash P_\lambda : I_\lambda \to O_\lambda\big\}_{\lambda \in \nat}$ and $\big\{\Delta_\lambda \vdash Q_\lambda : I_\lambda \to O_\lambda\big\}_{\lambda \in \nat}$ with identical typing judgments;

\item PPT family of interpretations $\big\{\int{-}_\lambda\big\}_{\lambda \in \nat}$ for $\Sigma$; and

\item an asymptotic theory that is sound with respect to $\big\{\int{-}_\lambda\big\}_{\lambda \in \nat}$;
\end{itemize}
we have that
\[ \mathlarger{\mathlarger{\vdash}} \; \big\{\Delta_\lambda \vdash P_\lambda \approx Q_\lambda : I_\lambda \to O_\lambda\big\}_{\lambda \in \nat}
\]
implies
\[\mathlarger{\mathlarger{\vDash}} \; \big\{\Delta_\lambda \vdash P_\lambda \approx Q_\lambda : I_\lambda \to O_\lambda\big\}_{\lambda \in \nat}.\]
\end{theorem}

\section{Case Studies}\label{sec:case_studies}
\renewcommand{\flip}{\mathsf{flip}}
\newcommand{\circuit}{\mathsf{circuit}}
\newcommand{\inputgate}{\textit{input-gate}}
\newcommand{\xorgate}{\textit{xor-gate}}
\newcommand{\andgate}{\textit{and-gate}}
\newcommand{\notgate}{\textit{not-gate}}
\newcommand{\party}{\mathsf{party}}
\newcommand{\ot}{\mathsf{ot}}
\newcommand{\Wires}{\mathsf{Wires}}
\newcommand{\Init}{\mathsf{Init}}
\newcommand{\Circ}{\mathsf{Circ}}
\newcommand{\Fin}{\mathsf{Fin}}
\newcommand{\Shares}{\mathsf{Shares}}
\newcommand{\OT}{\mathsf{1OutOf4OT}}
\renewcommand{\Adv}{\mathsf{Adv}}
\newcommand{\Ctrbs}{\mathsf{Ctrbs}}
\renewcommand{\Sim}{\mathsf{Sim}}
\newcommand{\Wire}{\mathsf{Wire}}
\newcommand{\SendInShare}{\mathsf{SendInShare}}
\newcommand{\SendOutShare}{\mathsf{SendOutShare}}
\newcommand{\InShareGen}{\mathsf{InShare}\textsf{-}\$}
\newcommand{\SumInShareGen}{\mathsf{InShare}\textsf{-}\$\textsf{-}\Sigma}
\newcommand{\InShare}{\mathsf{InShare}}
\newcommand{\SendBit}{\mathsf{SendBit}}
\newcommand{\RcvdBit}{\mathsf{RcvdBit}}
\newcommand{\Ctrb}{\mathsf{Ctrb}}
\newcommand{\SumCtrb}{\mathsf{Ctrb}\textsf{-}\Sigma}
\newcommand{\Share}{\mathsf{Share}}
\newcommand{\OutShare}{\mathsf{OutShare}}
\newcommand{\SumInShare}{\mathsf{InShare}\textsf{-}\Sigma}
\newcommand{\SumShare}{\mathsf{Share}\textsf{-}\Sigma}
\newcommand{\SumOutShare}{\mathsf{OutShare}\textsf{-}\Sigma}
\newcommand{\OTMsg}{\mathsf{OTMsg}}
\newcommand{\OTChc}{\mathsf{OTChc}}
\newcommand{\OTOut}{\mathsf{OTOut}}
\newcommand{\Col}{\mathsf{Col}}
\newcommand{\Row}{\mathsf{Row}}
\newcommand{\SumCol}{\mathsf{Col}\textsf{-}\Sigma}
\newcommand{\SumRow}{\mathsf{Row}\textsf{-}\Sigma}
\newcommand{\Sqr}{\mathsf{Sqr}}
\newcommand{\DiagRefl}{\mathsf{DiagRefl}}
\newcommand{\RowCol}{\mathsf{RowCol}}
\newcommand{\LeakInRcvd}{\mathsf{InRcvd}}
\newcommand{\LeakIn}{\mathsf{In}}
\newcommand{\LeakOut}{\mathsf{Out}}
\newcommand{\LeakRcvdInShare}{\mathsf{RcvdInShare}}
\newcommand{\LeakSendInShare}{\mathsf{SendInShare}}
\newcommand{\LeakRcvdOutShare}{\mathsf{RcvdOutShare}}
\newcommand{\LeakSendOutShare}{\mathsf{SendOutShare}}
\newcommand{\LeakInShareGen}{\mathsf{InShare}\textsf{-}\$}
\newcommand{\LeakSumInShareGen}{\mathsf{InShare}\textsf{-}\$\textsf{-}\Sigma}
\newcommand{\LeakInShare}{\mathsf{InShare}}
\newcommand{\LeakSendBit}{\mathsf{SendBit}}
\newcommand{\LeakRcvdBit}{\mathsf{RcvdBit}}
\newcommand{\LeakCtrb}{\mathsf{Ctrb}}
\newcommand{\LeakSumCtrb}{\mathsf{Ctrb}\textsf{-}\Sigma}
\newcommand{\LeakShare}{\mathsf{Share}}
\newcommand{\LeakOutShare}{\mathsf{OutShare}}
\newcommand{\LeakSumOutShare}{\mathsf{OutShare}\textsf{-}\Sigma}
\newcommand{\LeakOTMsg}{\mathsf{OTMsg}}
\newcommand{\LeakOTMsgRcvd}{\mathsf{OTMsgRcvd}}
\newcommand{\LeakOTChc}{\mathsf{OTChc}}
\newcommand{\LeakOTOut}{\mathsf{OTOut}}
\newcommand{\LeakOTChcRcvd}{\mathsf{OTChcRcvd}}
\newcommand{\InitOk}{\mathsf{Init}\textsf{-}\checkmark}
\newcommand{\WiresOk}{\mathsf{Wires}\textsf{-}\checkmark}
\newcommand{\SharesOk}{\mathsf{Shares}\textsf{-}\checkmark}
\newcommand{\CtrbsOk}{\mathsf{Ctrbs}\textsf{-}\checkmark}
\newcommand{\InOk}{\mathsf{In}\textsf{-}\checkmark}
\newcommand{\WireOk}{\mathsf{Wire}\textsf{-}\checkmark}
\newcommand{\InShareGenOk}{\mathsf{InShare}\textsf{-}\$\textsf{-}\checkmark}
\newcommand{\SumInShareGenOk}{\mathsf{InShare}\textsf{-}\$\textsf{-}\Sigma\textsf{-}\checkmark}
\newcommand{\InShareOk}{\mathsf{InShare}\textsf{-}\checkmark}
\newcommand{\SendBitOk}{\mathsf{SendBit}\textsf{-}\checkmark}
\newcommand{\RcvdBitOk}{\mathsf{RcvdBit}\textsf{-}\checkmark}
\newcommand{\CtrbOk}{\mathsf{Ctrb}\textsf{-}\checkmark}
\newcommand{\SumCtrbOk}{\mathsf{Ctrb}\textsf{-}\Sigma\textsf{-}\checkmark}
\newcommand{\ShareOk}{\mathsf{Share}\textsf{-}\checkmark}
\newcommand{\SumShareOk}{\mathsf{Share}\textsf{-}\Sigma\textsf{-}\checkmark}

We now briefly describe our four case studies, focusing on the Multi-Party GWM protocol. In Fig.~\ref{fig:case-studies}, we summarize our case studies by lines of code (second number in columns 2 and 3) and runtime to verify (column 4). Whenever applicable, we compare against the corresponding case study from \cite{ipdl} implemented in Coq (first number in columns 2 and 3). Our largest case study currently takes 12 minutes to verify. This number can likely be reduced with  performance optimizations and/or by splitting the proof across multiple files that can be verified in parallel\footnote{The case studies and the sources can be found at \url{https://github.com/concrete-bounds-for-mpc-proofs/concrete-bounds-for-mpc-simulation-proofs}.}.

\begin{figure}[ht]
  \begin{tabular}{|l|c|c|c|c|c|}
    \hline
    \multirow{2}{*}{Case Study} &
      \multicolumn{2}{c}{Proofs (LoC)} &
      \multicolumn{2}{c}{Defs (LoC)} &
      Runtime \\
    & Coq & DSL & Coq & DSL & DSL \\
    \hline
    Auth-To-Secure Channel \cite{constructive-crypto} & 128  & 61 & 97  & 143 &
      1.38s\\
    \hline
 DHKE-OTP \cite{advcompl} & 532 & 164 & 183 & 371 & 3.3s \\
 \hline
 Multi-Party Coin Flip \cite{blum83}& 1905 & 256 & 114 & 311 & 5.1s \\
 \hline
 Multi-Party GMW \cite{gmw}& - & 9102 & - & 5738 & 12.08m \\
 \hline
  \end{tabular}
\caption{Case Studies: size of proofs and definitions, and runtime to verify.}
\label{fig:case-studies}
\end{figure}

\paragraph{Authenticated-To-Secure Channel}

Our smallest case study constructs a secure channel from an authenticated one. Alice wants to communicate $q$ messages to Bob using an authenticated channel. The authenticated channel is not secure: it leaks each message to the adversary, and waits to receive an \textit{ok} message back from the adversary before delivering the in-flight message. Thus, the adversary cannot modify any of the messages but can read and delay them for any amount of time. To transmit information securely, Alice sends encryptions of her messages, which Bob decrypts using a shared key not known to the adversary. The indistinguishability assumption states that the encryption scheme is CPA-secure.

\paragraph{One-Time Pad (OTP) From Diffie-Hellman Key Exchange}

In symmetric-key encryption, the sender (Alice) and the receiver (Bob) need to agree on a shared secret key. One such key-agreement protocol is the Diffie-Hellman Key Exchange (DHKE), which assumes a cyclic group $G$ of a prime order with generator $g$. The indistinguishability assumption is the \emph{decisional Diffie-Hellman (DDH)} assumption: as long as the exponents $k,l$ are generated uniformly, even if the adversary knows the values $g^k$ and $g^l$, they will be unable to distinguish $(g^k)^l$ from a uniformly generated element of $G$.

We subsequently use the DHKE protocol to turn an authenticated channel into a one-time pad (OTP) that delivers a single secret message from Alice to Bob. Our proof is modular: in the first step, we establish that the DHKE protocol can be replaced with its idealization. In the second step, we prove that the resulting OTP protocol itself reduces to an idealization.

\paragraph{Multi-Party Coin Flip}

We implement a protocol (originally due to \cite{blum}) where $N+2$ parties labeled $0,\ldots,N-1$ reach a Boolean consensus. At the start of the protocol, each party commits to a randomly-generated Boolean. After all parties have committed, each party opens its commit. The consensus is the Boolean sum of all commits. We prove the protocol secure against a malicious attacker in the case when party $N$ is corrupt, party $N+1$ is honest, and any other party is arbitrarily honest or corrupt. As this protocol is perfectly secure, there are no indistinguishability assumptions.

\paragraph{Multi-Party GMW Protocol}

In the multi-party GMW protocol~\cite{gmw}, $N+2$ parties securely compute the value of a given Boolean circuit built out of \emph{xor-}, \emph{and-}, and \emph{not} gates. The inputs to the circuit are divided among the parties, and no party has access to the inputs of any other. Each party maintains its share of the value $v$ computed by each gate, and summing up the shares of all parties yields $v$. We prove the protocol secure in the case when party $N$ is semi-honest, party $N+1$ is honest, and any other party is arbitrarily honest or semi-honest. 

The protocol consists of the $N+2$ parties, plus an instance of the 1-Out-Of-4 Oblivious Transfer (OT) protocol for each gate and each pair of parties $n < m$, where $n$ is the sender and $m$ is the receiver. The code for each party is separated into three parts: in the initial phase, each party computes and distributes everyone’s shares for each of its inputs. In the inductive phase, each party computes their share of each gate by induction on the ambient circuit. At last, in the final phase, parties send their shares of each output wire to one another and add them up to compute the result.

The shares of each gate are computed as follows. In the case of an \emph{input} gate, the parties use the corresponding input share from the initial phase. In the case of a \emph{not} gate, parties $0,\ldots,N$ simply copy their share of the incoming wire, whereas party $N+1$ negates its share. If the gate is an \emph{xor} gate, the resulting share is the sum of the shares of the incoming two wires. The case of an \emph{and} gate is the most complex. The sum of everybody's shares must equal $\big(x_0 \oplus \ldots \oplus x_{N+1}\big) * \big(y_0 \oplus \ldots \oplus y_{N+1}\big)$, where $x_n,y_n$ are the respective shares of party $n$ on the incoming two wires. We have
\[\big(x_0 \oplus \ldots \oplus x_{N+1}\big) * \big(y_0 \oplus \ldots \oplus y_{N+1}\big) = \bigoplus_i \bigoplus_j x_i * y_j\]

\noindent Parties $n$ and $m$ engage in a 1-Out-Of-4 OT exchange to compute $(x_n * y_m) \oplus (x_m * y_n)$. There are four possible combinations of values that $x_m,y_m$ can take, and party $n$ computes the value of $(x_n * y_m) \oplus (x_m * y_n)$ for each. This offers party $m$ four messages to choose from, and he selects the one corresponding to the actual values of $x_m,y_m$. A small caveat: in the exchange as described above, party $m$ would still be able to infer the value of party $n$'s shares in certain cases: \emph{e.g.}, if $x_m = 0$ and $y_m = 1$, party $m$ gets the share $x_n$ as the result of the exchange. To prevent this, party $n$ encodes her messages by masking them with a random Boolean $b$ that only she knows. To offset for the presence of this Boolean, she includes it in her own share $b \oplus (x_n * y_n)$.

We assume four opaque protocols representing four implementations of the 1-Out-Of-4 Oblivious Transfer, one for each of the possible combinations of an honest/semi-honest sender and receiver. We also assume four cryptographic hardness axioms, stating that each OT implementation is approximately equal to an ideal 1-Out-Of-4 OT functionality. In the first step of our proof, we replace each OT implementation by its the corresponding functionality. The rest of the proof is carried out in the exact fragment of our proof system.

Since the last party is by assumption honest, the simulator does not have access to its inputs. Therefore, any computation that depends on the value of the inputs belonging to the last party must be eliminated. In particular, all shares of the last party must be eliminated. Instead, the simulator only computes shares for parties $0,\ldots,N$ in the inductive part, and replaces every mention of the last party's share in its leakage by the quantity $x \oplus \big(\bigoplus_{i \leq N} x_i\big)$. Here $x$ is the value carried by the gate as computed by the ideal functionality, and leaked to the simulator by a semi-honest party.

In this case study, the main challenge for formal verification is how to effectively carry out proofs by induction nested three levels deep: one for the arbitrary Boolean circuit, and two for each of the arbitrarily many parties communicating with another party. Additionally, we have to account for an essentially arbitrary combination of honest/semi-honest parties, as all but the last two parties can be honest or semi-honest. Every one of these factors reflects into the actual code of the protocol, and has to be taken into account when computing concrete security bounds. The final concrete bound computed by our tool is as follows:

\begin{lstlisting}
indistinguishability assumption HH2HH :
count: |{(n, m, k) s.t. n < N + 2, m < N + 2, k < K,
  when isHonest(n) and isHonest(m)}|
context:
  N * K * max(N * 434 + 990, N * 434 + 1011, N * 607 + 1329, N * 434 + 983) +
  N * N * K * max(|1OutOf4OTReal-Honest-Honest|, 543) + N * N * K * 218 +
  N * K * max(|1OutOf4OTReal-Honest-Honest|, 543) * 4 + N * K * 932 +
  K * max(N * 434 + 990, N * 434 + 1011, N * 607 + 1329, N * 434 + 983) * 2 +
  N * N * maxValue(I, N + 2) * 300 +  N * N * 68 + K * 992 + N * 352 +
  K * max(|1OutOf4OTReal-Honest-Honest|, 543) * 3 +
  (K - 1) * max(|1OutOf4OTReal-Honest-Honest|, 543) +
  N * maxValue(I, N + 2) * 1169 + maxValue(I, N + 2) * 1138 + 509

indistinguishability assumption SHH2SHH : ...
indistinguishability assumption HSH2HSH : ...
indistinguishability assumption SHSH2SHSH : ...
\end{lstlisting}
Here \code{K} is the number of wires in the circuit, \code{HH2HH} is the indistinguishability assumption stating that the 1-Out-Of-4 Oblivious Transfer protocol when both the sender and the receiver are honest can be soundly replaced by the corresponding functionality, and \code{maxValue(I, N + 2)} selects the maximum of \code{I(0),...,I(N + 1)}, where \code{I(n)} denotes the number of inputs to party \code{n}. We omit the bounds for the other indistinguishability assumptions, as these are entirely analogous. As expected, the bound \code{context} is $\mathsf{O}(\text{\code{N * N * K}})$, since we have one instance of an OT protocol for each of the $K$ wires and each pair of parties. As we can see from the the bound \code{count}, the total number of times we invoke an indistinguishability assumption is \code{(N + 2) * (N + 2) * K}, since we replace each of the OT protocols with its idealization. The indistinguishability assumption applicable to a particular OT instance is conditional on whether the sender and receiver are honest or semi-honest, respectively.

\paragraph{DSL Implementation}

Our DSL serves as a layer of abstraction over the implementation internals, hiding low-level proof details  from the user. For each rule, we have language construct in the DSL that handles its application. For example, an application of the rule \textsc{subst} in Fig.~\ref{fig:protocols_equality_strict} is written in the DSL as \code{subst o1 into o2}. Internally, this translates into a call of a Maude strategy, which applies the substitution rule over a protocol without explicitly writing applications of congruence or exchange rules. Moreover, in many cases these strategies are generated on the fly, instead of requiring the user to manually specify them.

\section{Future Work}
We present a promising formalism for developing verified cryptographic proofs for MPC protocols. We compute concrete security bounds by bounding the size of the program context in which indistinguishability assumptions are applied, and analyzing the cost of the interpreter that executes the program context. Currently, our main source of over-approximation is the runtime of this interpreter, which is currently quadratic due to using linear scans for inputs (see Section~\ref{sec:absorption}); however, reducing the overhead to near-linear is possible using standard techniques.

Our tool can model arbitrary (static) corruption scenarios, including protocols that tolerate $k/N$ corruptions (either semi-honest or malicious); formalizing a protocol that demonstrates this capability is one avenue for future work. In in addition to verifying even more sophisticated MPC protocols (\emph{e.g.}, garbled circuits), we want to develop new algorithms for automatically \emph{synthesizing} cryptographic simulators, rather than requiring them to be manually encoded. Finally, we aim to extend our system to handle more expressive classes of protocols, such as those exhibiting threshold behavior (\emph{e.g.}, consensus protocols).

\bibliography{IEEEabrv,main}

\appendix
%\section{Appendix}
\section{Turing Machine Bounds}\label{sec:tm_bound}

The Turing Machine bound of a type $\tau$ is straightforward:
\begin{align*}
\tmnorm{\type_i} & \coloneqq t_i \\
\tmnorm{\one} & \coloneqq 0 \\
\tmnorm{\Bool} & \coloneqq 1 \\
\tmnorm{\tau_1 \times \tau_2} & \coloneqq \tmnorm{\tau_1} + \tmnorm{\tau_2}
\end{align*}
The encoding of variables $x$ of type $\tau$ uses the symbols \textsf{``$\mathsf{(}$''},
\textsf{``$\mathsf{var}$''}, \textsf{``$\mathsf{:}$''}, \textsf{``$\mathsf{)}$''} in addition to the de Bruijn index of the variable $x$, encoded as a single symbol, and the encoding of the type annotation $\tau$. For expressions $\checkmark$, $\true$, $\false$, we use the corresponding symbols \textsf{``$\checkmark$''}, \textsf{``$\mathsf{true}$''}, \textsf{``$\mathsf{false}$''} and the two parenthesis symbols \textsf{``$\mathsf{(}$''}, \textsf{``$\mathsf{)}$''}. For an application $\func \ e$ of a function of type $\sigma \to \tau$, we use the symbols \textsf{``$\mathsf{(}$''}, \textsf{``$\mathsf{app}$''}, \textsf{``$\to$''}, \textsf{``$\mathsf{)}$''} in addition to the function symbol $\func$, encoded as a single symbol, and the encodings of the two type annotations $\sigma, \tau$ and the expression $e$. To encode a pair $(e_1, e_2)$, we will only need the encodings of the two expressions $e_1$ and $e_2$. Finally, to encode first and second projections, we will use the symbols \textsf{``$\mathsf{(}$''}, \textsf{``$\mathsf{fst}$''}, \textsf{``$\mathsf{snd}$''}, \textsf{``$\times$''}, \textsf{``$\mathsf{of}$''}, \textsf{$\mathsf{)}$''} in addition to the encodings of the two type annotations $\sigma, \tau$ and the expression $e$.
\begin{align*}
\tmnorm{\Var{x}{\tau}} & \coloneqq \tmnorm{\tau} + 5 \\
\tmnorm{\checkmark} & \coloneqq 3 \\
\tmnorm{\true} & \coloneqq 3 \\
\tmnorm{\false} & \coloneqq 3 \\
\tmnorm{\App{\func}{\sigma}{\tau}{e}} & \coloneqq \tmnorm{\sigma} + \tmnorm{\tau} + \tmnorm{e} + 5 \\
\tmnorm{(e_1, e_2)} & \coloneqq \tmnorm{e_1} + \tmnorm{e_2} \\
\tmnorm{\fst_{\sigma \times \tau} \ e} & \coloneqq \tmnorm{\sigma} + \tmnorm{\tau} + \tmnorm{e} + 5 \\
\tmnorm{\snd_{\sigma \times \tau} \ e} & \coloneqq \tmnorm{\sigma} + \tmnorm{\tau} + \tmnorm{e} + 5
\end{align*}
For a $\ret{e}$, we use the symbols \textsf{``$\mathsf{(}$''}, \textsf{``$\mathsf{ret}$''}, \textsf{$\mathsf{)}$''} in addition to the encoding of the expression $e$. For a sampling $\samp{\dist}{e}$ from a distribution of type $\sigma \to \tau$, we use the symbols \textsf{``$\mathsf{(}$''}, \textsf{``$\mathsf{samp}$''}, \textsf{``$\twoheadrightarrow$''}, \textsf{``$\mathsf{)}$''} in addition to the distribution symbol $\dist$, encoded as a single symbol, and the encodings of the two type annotations $\sigma, \tau$ and the expression $e$. For a $\read{c}$ from a channel of type $\tau$, we use the symbols \textsf{``$\mathsf{(}$''}, \textsf{``$\mathsf{read}$''}, \textsf{``$\mathsf{:}$''}, \textsf{``$\mathsf{)}$''} in addition to the de Bruijn index of the channel $c$, encoded as a single symbol, and the encoding of the type annotation $\tau$. Furthermore, we will need one extra symbol: one of \textsf{``$\mathsf{input}$-$\mathsf{to}$-$\mathsf{query}$''}, \textsf{``$\mathsf{input}$-$\mathsf{queried}$''}, \textsf{``$\mathsf{input}$-$\mathsf{not}$-$\mathsf{to}$-$\mathsf{query}$''}. When encoding a protocol $Q : I \cup O_1 \to O_2$ coming from the \textsc{comp-cong} rule, we use \textsf{``$\mathsf{input}$-$\mathsf{to}$-$\mathsf{query}$''} or \textsf{``$\mathsf{input}$-$\mathsf{queried}$''} if we are reading from a channel $o_1 \in O_1$, according to whether we have already queried the channel $o_1$, and \textsf{``$\mathsf{input}$-$\mathsf{not}$-$\mathsf{to}$-$\mathsf{query}$''} otherwise.

For a conditional $\ifte{e}{R_1}{R_2}$, we use the symbols \textsf{``$\mathsf{(}$''}, \textsf{``$\mathsf{if}$''}, \textsf{``$\mathsf{then}$''}, \textsf{``$\mathsf{else}$''}, \textsf{``$\mathsf{)}$''} in addition to the encodings of the expression $e$ and the two reactions $R_1, R_2$. Finally, to encode a bind, we use the symbols \textsf{``$\{$''}, \textsf{``$\_$''}, \textsf{``$\mathsf{:}$''}, \textsf{``$\leftarrow$''}, \textsf{``$\mathsf{;}$''}, \textsf{``$\}$''} in addition to the encodings of the type annotation $\sigma$ and the two reactions $R$ and $S$. The symbol \textsf{``$\_$''} is used in lieu of the bound variable name $x$ and stands for de Bruijn index $0$.
\begin{align*}
\tmnorm{\ret{e}} & \coloneqq \tmnorm{e} + 3 \\
\tmnorm{\Samp{\dist}{\sigma}{\tau}{e}} & \coloneqq \tmnorm{\sigma} + \tmnorm{\tau} + \tmnorm{e} + 5 \\
\tmnorm{\Read{c}{\tau}} & \coloneqq \tmnorm{\tau} + 6 \\
\tmnorm{\ifte{e}{R_1}{R_2}} & \coloneqq \tmnorm{e} + \tmnorm{R_1} + \tmnorm{R_2} + 5 \\
\tmnorm{x : \sigma \leftarrow R; \ S} & \coloneqq \tmnorm{\sigma} + \tmnorm{R} + \tmnorm{S} + 6
\end{align*}
To encode the zero protocol $\zero$, we use the single symbol \textsf{``$\mathsf{0}$''}. For an assignment $\assign{o}{R}$, we use the symbols \textsf{``$\mathsf{[}$''}, \textsf{``$\coloneqq$''}, \textsf{``$\mathsf{react}$''}, \textsf{``$\mathsf{]}$''} in addition to the de Bruijn index of the channel $c$, encoded as a single symbol, and the encoding of the reaction $R$. For a parallel composition $\Par{P}{Q}$, we use the symbols \textsf{``$\mathsf{(}$''}, \textsf{``$\|$''}, \textsf{``$\mathsf{)}$''} in addition to the encodings of the two protocols $P$ and $Q$. Finally, for the declaration of a new channel $\new{o}{\tau}{P}$, we use the symbols \textsf{``$\mathsf{new}$''}, \textsf{``$\_$''}, \textsf{``$\mathsf{:}$''}, \textsf{``$\mathsf{in}$''}, \textsf{``$\mathsf{wen}$''} in addition to the encodings of the typing annotation $\tau$ and the protocol $P$. The symbol \textsf{``$\_$''} is used in lieu of the bound channel name $c$ and stands for de Bruijn index $0$. The symbol \textsf{``$\mathsf{wen}$''} indicates the end of the binding scope.
\begin{align*}
\tmnorm{\zero} & \coloneqq 1 \\
\tmnorm{\assign{o}{R}} & \coloneqq \tmnorm{R} + 5 \\
\tmnorm{\Par{P}{Q}} & \coloneqq \tmnorm{P} + \tmnorm{Q} + 3 \\
\tmnorm{\new{c}{\tau}{P}} & \coloneqq \tmnorm{\tau} + \tmnorm{P} + 5
\end{align*}

%%%%%%%%%%%%%%%%%%%%%%%%%%%%%%%%%%%%%%%%%%%%%%%%%%%%%%%%%%%%%%%%%

\section{Encoding Protocols on a Turing Machine Tape}

To encode protocols on a Turing Machine tape, we make use of the following sets of symbols:
\begin{itemize}
\item $\mathsf{Punc}$ with symbols \textsf{``$\langle$''},\textsf{``$\rangle$''}, \textsf{``$($''}, \textsf{``$)$''}, \textsf{``$\{$''}, \textsf{``$\}$''}, \textsf{``$[$''}, \textsf{``$]$''}, \textsf{``$\_$''}, \textsf{``$:$''}, \textsf{``$\cdot$''}, \textsf{``$;$''}, \textsf{``$\to$''}, \textsf{``$\twoheadrightarrow$''}, \textsf{``$\leftarrow$''}, \textsf{``$\times$''}, \textsf{``$\coloneqq$''}, \textsf{``$\|$''}, \textsf{``$\parm$''}, 

\item $\mathsf{KeyWords}$ with symbols \textsf{``$\mathsf{var}$''}, \textsf{``$\mathsf{\checkmark}$''}, \textsf{``$\mathsf{true}$''}, \textsf{``$\mathsf{false}$''}, \textsf{``$\mathsf{app}$''}, \textsf{
``$\mathsf{fst}$''}, \textsf{``$\mathsf{snd}$''}, \textsf{``$\mathsf{of}$''}, \textsf{``$\mathsf{ret}$''}, \textsf{``$\mathsf{samp}$''}, \textsf{``$\mathsf{read}$''}, \textsf{``$\mathsf{if}$''}, \textsf{``$\mathsf{then}$''}, \textsf{``$\mathsf{else}$''}, \textsf{``$\mathsf{0}$''}, \textsf{``$\mathsf{new}$''}, \textsf{``$\mathsf{in}$''}, \textsf{``$\mathsf{wen}$''}, \textsf{``$\mathsf{input}$-$\mathsf{to}$-$\mathsf{query}$''}, \textsf{``$\mathsf{input}$-$\mathsf{queried}$''}, \textsf{``$\mathsf{input}$-$\mathsf{not}$-$\mathsf{to}$-$\mathsf{query}$''}.
\end{itemize}

\noindent We also need a finite set of de Bruijn indices in lieu of channel and variable names. To derive an upper bound on how many indices we will need, we statically count the maximum depth of variable and channel declarations in the protocol, giving us a \emph{variable-index bound} and a \emph{channel-index bound}, respectively.

To avoid an infinite loop, an adversary executing the absorbed protocol will need to keep track of which channels have already been queried for a value. We store this information inside the protocol in the form of an annotation: for each channel read $\Read{c}{\tau}$, we denote whether the channel $c$ has already been queried for a value, if applicable. By erasing the annotations from a query-annotated reaction or protocol, we obtain the underlying IPDL construct.

Given an ambient interpretation $\int{-}$ for the signature $\Sigma$, we now show how to encode IPDL constructs as a sequence of symbols on a Turing Machine tape. For types, the encoding $\tmenc{\tau}$ consists of the symbol \textsf{``$\cdot$''} repeated $|\tau|$-many times. For expressions, we have the encoding below, where + denotes string concatenation. We recall that each variable name is represented as a de Bruijn index, and is in particular a natural number.
\begin{align*}
\tmenc{v} & \coloneqq v \\
\tmenc{\Var{x}{\tau}} & \coloneqq \textsf{``$($''} + \textsf{``$\mathsf{var}$''} + x + \textsf{``$:$''} + \tmenc{\tau} + \textsf{``$\mathsf{)}$''} \\
\tmenc{\checkmark} & \coloneqq \textsf{``$($''} + \textsf{``$\checkmark$''} + \textsf{``$)$''} \\
\tmenc{\true} & \coloneqq \textsf{``$($''} + \textsf{``$\mathsf{true}$''} + \textsf{``$)$''} \\
\tmenc{\false} & \coloneqq \textsf{``$($''} + \textsf{``$\mathsf{false}$''} + \textsf{``$)$''} \\
\tmenc{\App{\func}{\sigma}{\tau}{e}} & \coloneqq \textsf{``$($''} + \textsf{``$\mathsf{app}$''} + \tmenc{\sigma} + \textsf{``$\to$''} + \tmenc{\tau} + \func + \tmenc{e} + \textsf{``$)$''} \\
\tmenc{(e_1, e_2)} & \coloneqq \tmenc{e_1} + \tmenc{e_2} \\
\tmenc{\fst_{\sigma \times \tau} \ e} & \coloneqq \textsf{``$($''} + \textsf{``$\mathsf{fst}$''} + \tmenc{\sigma} + \textsf{``$\times$''} + \tmenc{\tau} + \textsf{``$\mathsf{of}$''} + \tmenc{e} + \textsf{``$)$''} \\
\tmenc{\snd_{\sigma \times \tau} \ e} & \coloneqq \textsf{``$($''} + \textsf{``$\mathsf{snd}$''} + \tmenc{\sigma} + \textsf{``$\times$''} + \tmenc{\tau} + \textsf{``$\mathsf{of}$''} + \tmenc{e} + \textsf{``$)$''}
\end{align*}
The encoding $\tmenc{a}$ of an annotation is the corresponding symbol. For reactions, we have the following encoding, where we recall that each channel name is represented as a de Bruijn index, and is in particular a natural number.
\begin{align*}
\tmenc{\val{v}} & \coloneqq \textsf{``$\langle$''} + v + \textsf{``$\rangle$''} \\
\tmenc{\ret{e}} & \coloneqq \textsf{``$($''} + \textsf{``$\mathsf{ret}$''} + \tmenc{e} + \textsf{``$)$''} \\
\tmenc{\Samp{\dist}{\sigma}{\tau}{e}} & \coloneqq \textsf{``$($''} + \textsf{``$\mathsf{samp}$''} + \tmenc{\sigma} + \textsf{``$\twoheadrightarrow$''} + \tmenc{\tau} + \dist + \tmenc{e} + \textsf{``$)$''} \\
\tmenc{\QAnnRead{c}{\tau}{a}} & \coloneqq \textsf{``$($''} + \textsf{``$\mathsf{read}$''} + \tmenc{a} + c + \textsf{``$:$''} + \tmenc{\tau} + \textsf{``$)$''} \\
\tmenc{\ifte{e}{R_1}{R_2}} & \coloneqq \textsf{``$($''} + \textsf{``$\mathsf{if}$''} + \tmenc{e} + \textsf{``$\mathsf{then}$''} + \tmenc{R_1} + \textsf{``$\mathsf{else}$''} + \tmenc{R_2} + \textsf{``$)$''} \\
\tmenc{x : \sigma \leftarrow R; \ S} & \coloneqq \textsf{``$\{$''} + \textsf{``$\_$''} + \textsf{``:''} + \tmenc{\sigma} + \textsf{``$\leftarrow$''} + \tmenc{R} + \textsf{``;''} + \tmenc{S} + \textsf{``$\}$''}
\end{align*}
Finally, for protocols we have the encoding below.
\begin{align*}
\tmenc{\zero} & \coloneqq \textsf{``$\mathsf{0}$''} \\
\tmenc{\assign{o}{v}} & \coloneqq \textsf{``$[$''} + o + \textsf{``$\coloneqq$''} + v + \textsf{``$]$''} \\
\tmenc{\assign{o}{R}} & \coloneqq \textsf{``$($''} + o + \textsf{``$\coloneqq$''} + \textsf{``$\mathsf{react}$''} + \tmenc{R} + \textsf{``$)$} \\
\tmenc{\Par{P}{Q}} & \coloneqq \textsf{``$($''} + \tmenc{P} + \textsf{``$\|$''} + \tmenc{Q} + \textsf{``$)$''} \\
\tmenc{\new{c}{\tau}{P}} & \coloneqq \textsf{``$\mathsf{new}$''} + \textsf{``$\_$''} + \textsf{``$:$''} + \tmenc{\tau} + \textsf{``$\mathsf{in}$''} + \tmenc{P} + \textsf{``$\mathsf{wen}$''}
\end{align*}
To avoid having to shift the tape contents when executing IPDL protocols on a Turing Machine tape, we will make use of the white-space symbol \textsf{`` ''}, which we consider as distinct from the symbol \emph{blank}. The former will be used as a placeholder so that our protocol encoding remains at a constant length throughout the execution. For this reason, we extend our notion of encoding to allow extra white-spaces around the encoding of an expression $e$ or a query-annotated reaction $R$ occurring inside a query-annotated protocol $P$.

%%%%%%%%%%%%%%%%%%%%%%%%%%%%%%%%%%%%%%%%%%%%%%%%%%%%%%%%%%%%%%%%%

\section{Layered Approximate Judgements}\label{sec:layered}

The layered form of our approximate judgments is shown in Figures~\ref{fig:protocols_layered_1} and~\ref{fig:protocols_layered_2}.

\begin{figure*}
\begin{mathpar}
\fbox{$\Delta \vdash \lapproxcong{P}{Q}{I}{O}{k}{\psi}{0}$}\\
\inferrule*[right=axiom]{\Delta^k \vdash P^k \approx Q^k : I^k \to O^k}{\Delta^k \vdash \lapproxcong{P^k}{Q^k}{I^k}{O^k}{k}{0}{0}}\\
\fbox{$\Delta \vdash \lapproxcong{P}{Q}{I}{O}{k}{\psi}{1}$}\\
\inferrule*[right=sub]{\Delta \vdash \lapproxcong{P}{Q}{I}{O}{k}{\psi}{0}}{\Delta \vdash \lapproxcong{P}{Q}{I}{O}{k}{\psi}{1}}\and\inferrule*[right=embed]{\phi : \Delta_1 \to \Delta_2 \\ \Delta_2 \vdash \lapproxcong{P}{Q}{I}{O}{k}{\psi}{0}}{\Delta_1 \vdash \lapproxcong{\phi^\star(P)}{\phi^\star(Q)}{\phi^\star(I)}{\phi^\star(O)}{k}{\psi}{1}}\\
\fbox{$\Delta \vdash \lapproxcong{P}{Q}{I}{O}{k}{\psi}{2}$}\\
\inferrule*[right=sub]{\Delta \vdash \lapproxcong{P}{Q}{I}{O}{k}{\psi}{1}}{\Delta \vdash \lapproxcong{P}{Q}{I}{O}{k}{\psi}{2}}\and
\inferrule*[right=input-unused]{i \notin I \cup O \\ \Delta \vdash \lapproxcong{P}{Q}{I}{O}{k}{\psi}{2}}{\Delta \vdash \lapproxcong{P}{Q}{I \cup \{i\}}{O}{k}{\psi}{2}}\\
\fbox{$\Delta \vdash \lapproxcong{P}{Q}{I}{O}{k}{\psi}{3}$}\\
\inferrule*[right=sub]{\Delta \vdash \lapproxcong{P}{Q}{I}{O}{k}{\psi}{2}}{\Delta \vdash \lapproxcong{P}{Q}{I}{O}{k}{\psi}{3}}\and
\inferrule*[right=cong-comp-left]{\Delta \vdash \lapproxcong{P}{P'}{I \cup O_2}{O_1}{k}{\psi}{2} \\ \Delta \vdash Q : I \cup O_1 \to O_2}{\Delta \vdash \lapproxcong{\Par{P}{Q}}{\Par{P'}{Q}}{I}{O_1 \cup O_2}{k}{\psi + \tmnorm{Q} + 3}{3}}\\
\fbox{$\Delta \vdash \lapproxcong{P}{Q}{I}{O}{k}{\psi}{4}$}\\
\inferrule*[right=sub]{\Delta \vdash \lapproxcong{P}{Q}{I}{O}{k}{\psi}{3}}{\Delta \vdash \lapproxcong{P}{Q}{I}{O}{k}{\psi}{4}}\and
\inferrule*[right=cong-new]{\Delta, o : \tau \vdash \lapproxcong{P}{P'}{I}{O \cup \{o\}}{k}{\psi}{4}}{\Delta \vdash \lapproxcong{\big(\new{o}{\tau}{P}\big)}{\big(\new{o}{\tau}{P'}\big)}{I}{O}{k}{\psi}{4}}
\end{mathpar}
\caption{Layered approximate judgements for protocols.}
\label{fig:protocols_layered_1}
\end{figure*}

\begin{figure*}
\begin{mathpar}
\fbox{$\Delta \vdash \lapproxcong{P}{Q}{I}{O}{k}{\psi}{5}$}\\
\inferrule*[right=sub]{\Delta \vdash P = P' : I \to O \\  \Delta \vdash \lapproxcong{P'}{Q'}{I}{O}{k}{\psi}{4} \\ \Delta \vdash Q' = Q : I \to O}{\Delta \vdash \lapproxcong{P}{Q}{I}{O}{k}{\psi}{5}}\\
\fbox{$\Delta \vdash \lapproxeq{P}{Q}{I}{O}{\xi}{\psi}{5}$}\\
\inferrule*[right=strict]{\Delta \vdash P = Q : I \to O}{\Delta \vdash \lapproxeq{P}{Q}{I}{O}{(i \mapsto 0)}{(i \mapsto 0)}{5}}\and
\inferrule*[right=approx-cong]{\Delta \vdash \lapproxcong{P}{Q}{I}{O}{k}{\psi}{5}}{\Delta \vdash \lapproxeq{P}{Q}{I}{O}{\big(k \mapsto 1, i \neq k \mapsto 0\big)}{\big(k \mapsto \psi, i \neq k \mapsto 0\big)}{5}}\and
\inferrule*[right=sym]{\Delta \vdash \lapproxeq{P_1}{P_2}{I}{O}{\xi}{\psi}{5}}{\Delta \vdash \lapproxeq{P_2}{P_1}{I}{O}{\xi}{\psi}{5}}\and
\inferrule*[right=trans]{\Delta \vdash \lapproxeq{P_1}{P_2}{I}{O}{\xi_1}{\psi_1}{5} \\ \Delta \vdash \lapproxeq{P_2}{P_3}{I}{O}{\xi_2}{\psi_2}{5}}{\Delta \vdash \lapproxeq{P_1}{P_3}{I}{O}{\big(i \mapsto \xi_1(i) + \xi_2(i)\big)}{\big(i \mapsto \max(\psi_1(i), \psi_2(i))\big)}{5}}
\end{mathpar}
\caption{Layered approximate judgements for protocols, continued.}
\label{fig:protocols_layered_2}
\end{figure*}

\begin{lemma}
We have $\Delta \vdash \approxeq{P}{Q}{I}{O}{k}{l}$ iff $\Delta \vdash \lapproxeq{P}{Q}{I}{O}{k}{l}{5}$ for any protocols $\Delta \vdash P : I \to O$ and $\Delta \vdash Q : I \to O$.
\end{lemma}

%%%%%%%%%%%%%%%%%%%%%%%%%%%%%%%%%%%%%%%%%%%%%%%%%%%%%%%%%%%%%%%%%

\section{Perfect Indistinguishability: Proof of Lemma~1}
\label{sec:proof-lemma1}

\begin{lemma*}[Perfect indistinguishability]
For any interpretation $\int{-}$ for $\Sigma$, derivation $\Delta \vdash P = Q : I \to O$, and adversary $\Adv$ for protocols $\Delta \vdash I \to O$, we have
\[\Big|\mathsf{Pr}\big[\interaction{\Adv}{P}{\int{-}} = 1\big] - \mathsf{Pr}\big[\interaction{\Adv}{Q}{\int{-}} = 1\big]\Big|= 0.\]
\end{lemma*}

\begin{proof}
Fix an adversary $\Adv$ as in Definition~3. By assumption, we have a proof $\Delta \vdash P = Q : I \to O$, which means we also have a proof that $\Delta' \vdash \phi^\star(P) = \phi^\star(Q) : \phi^\star(I) \to \phi^\star(O)$. The soundness theorem for strict equality of protocols applied to this proof gives us a bisimulation $\sim$ such that $1[\phi^\star(P)] \sim 1[\phi^\star(Q)]$. Now let $\sim_\adv$ be a binary relation on sub-distributions on pairs where the first element is an adversary state and the second is a protocol of type $\Delta \vdash I \to O$, defined as follows:
\begin{itemize}
\item $(s,\eta) \sim_\adv (s,\varepsilon)$ if $s \in \St$ and $\eta \sim \varepsilon$, where we use a distribution in place of a protocol to indicate the obvious lifting to sub-distributions on pairs of the the aforementioned form, and
\item $1[\bot] \sim_\adv 1[\bot]$, where $\bot$ indicates that the security game between the adversary and the protocol halted without a decision Boolean.
\end{itemize}
Let $\mathcal{L}_{\sim_\adv}$ be the closure of $\sim_\adv$ under joint convex combinations. Explicitly, $\mathcal{L}_{\sim_\adv}$ is defined by
\[\Big(\sum_i c_i \, \eta_i\Big) \; \mathcal{L}_{\sim_\adv} \; \Big(\sum_i c_i \, \varepsilon_i\Big)\]
for coefficients $c_i > 0$ with $\sum_i c_i = 1$ and distributions $\eta_i \sim_\adv \varepsilon_i$. We now establish a loop invariant for the algorithm in Figure~7. Before starting the first round, the initial distributions are suitably related: by assumption, we have $1[\phi^\star(P)] \sim 1[\phi^\star(Q)]$, which means that
\[1\big[(s_\star,\phi^\star(P))\big] \; \mathcal{L}_{\sim_\adv} \; 1\big[(s_\star,\phi^\star(Q))\big]\]
as the two distributions are already related under $\sim_\adv$. Now assume that we have two sub-distributions related by $\mathcal{L}_{\sim_\adv}$. We prove that performing a single round yields sub-distributions that are again related by $\mathcal{L}_{\sim_\adv}$. It suffices to show this for the case $(s,\eta) \sim_\adv (s,\varepsilon)$, where $s \in \St$ and $\eta \sim \varepsilon$. We first compute the distributions $\eta' \coloneqq \eval{\eta}$ and $\varepsilon' \coloneqq \eval{\varepsilon}$. By definition of $\sim$ we have $\eta' \sim \varepsilon'$. Independently, we probabilistically compute the type of interaction to perform together with a new adversary state $s'$. If no interaction has been chosen, the resulting distributions are $(s',\eta')$ and $(s',\varepsilon')$. We have
\[(s',\eta') \; \mathcal{L}_{\sim_\adv} \; (s',\varepsilon')\]
as desired, as the two distributions are already related under $\sim_\adv$. If the interaction is an input on channel $i$, we compute $\Out_i(s')$ to see if in the adversary's current state $s'$ the channel $i$ carries a value. If this is not the case, the resulting distributions are $(s',\eta')$ and $(s',\varepsilon')$. Here we again have $(s',\eta') \; \mathcal{L}_{\sim_\adv} \; (s',\varepsilon')$, as desired. On the other hand, if the channel $i$ carries a value $v$, the resulting distributions are $\big(s',\eta'[\read{i} := \val{v}]\big)$ and $\big(s',\varepsilon'[\read{i} := \val{v}]\big)$. Now because $\eta' \sim \varepsilon'$, by definition of $\sim$ we have $\eta'[\read{i} := \val{v}] \sim \varepsilon'[\read{i} := \val{v}]$. Thus we have
\[\big(s',\eta'[\read{i} := \val{v}]\big) \; \mathcal{L}_{\sim_\adv} \; \big(s',\varepsilon'[\read{i} := \val{v}]\big)\]
as desired, as the two distributions are already related under $\sim_\adv$. Finally, if the interaction is a query for an output channel $o$, we recall that the valuation property of the bisimulation $\sim$ allows us to jointly partition the distributions $\eta' \sim \varepsilon'$ into a joint convex combination \[\eta' = \sum_i c_i \, \eta'_i \; \sim \, \sum_i c_i \, \varepsilon'_i = \varepsilon'\]
with $c_i > 0$ and $\sum_i c_i = 1$ such that
\begin{itemize}
\item the respective components $\eta'_i \sim \varepsilon'_i$ are again related, and
\item $\valueat{\eta'_i}{o} = v_\bot = \valueat{\varepsilon'_i}{o}$ for the same $v_\bot \in \{\bot\} \cup \int{\tau}$ where $o : \tau$ in $\Delta'$.
\end{itemize}
Therefore, it suffices to consider the respective components $\eta'_i \sim \varepsilon'_i$ with the same $v_\bot$. If $v_\bot$ is $\bot$, then the resulting distributions are $(s',\eta'_i)$ and $(s',\varepsilon'_i)$. Here we again have $(s',\eta'_i) \; \mathcal{L}_{\sim_\adv} \; (s',\varepsilon'_i)$, as desired. On the other hand, if $v_\bot$ is a value $v$, then the resulting distributions are $\big(\In_o(v,s'),\eta'_i\big)$ and $\big(\In_o(v,s'),\varepsilon'_i\big)$. Thus we have
\[\big(\In_o(v,s'),\eta'_i\big) \; \mathcal{L}_{\sim_\adv} \; \big(\In_o(v,s'),\varepsilon'_i\big)\]
as desired, as the two distributions are already related under $\sim_\adv$. This proves that after completing the required number of rounds, we end up with two sub-distributions related by $\mathcal{L}_{\sim_\adv}$. It is now easy to see that they induce the same sub-distribution on decision Booleans. It suffices to prove this for the case $(s,\eta) \sim_\adv (s,\varepsilon)$, where $s \in \St$ and $\eta \sim \varepsilon$. But the state $s$ is the same for both distributions, so the resulting distribution on decision Booleans is $1[\Dec(s)]$. This finishes the proof.
\end{proof}

%%%%%%%%%%%%%%%%%%%%%%%%%%%%%%%%%%%%%%%%%%%%%%%%%%%%%%%%%%%%%%%%%

\section{Soundness of Asymptotic Equality: Proof of Theorem~2}\label{sec:soundness_asympto}

\begin{theorem*}[Soundness of asymptotic equality of protocols]
For any
\begin{itemize}
\item protocol families $\big\{\Delta_\lambda \vdash P_\lambda : I_\lambda \to O_\lambda\big\}_{\lambda \in \nat}$ and $\big\{\Delta_\lambda \vdash Q_\lambda : I_\lambda \to O_\lambda\big\}_{\lambda \in \nat}$ with identical typing judgments;

\item PPT family of interpretations $\big\{\int{-}_\lambda\big\}_{\lambda \in \nat}$ for $\Sigma$; and

\item an asymptotic theory that is sound with respect to $\big\{\int{-}_\lambda\big\}_{\lambda \in \nat}$;
\end{itemize}
we have that
\[ \mathlarger{\mathlarger{\vdash}} \; \big\{\Delta_\lambda \vdash P_\lambda \approx Q_\lambda : I_\lambda \to O_\lambda\big\}_{\lambda \in \nat}
\]
implies
\[\mathlarger{\mathlarger{\vDash}} \; \big\{\Delta_\lambda \vdash P_\lambda \approx Q_\lambda : I_\lambda \to O_\lambda\big\}_{\lambda \in \nat}.\]
\end{theorem*}

\begin{proof}
Given $n$ approximate axioms, the top-level asymptotic equality judgement
\[\mathlarger{\mathlarger{\vdash}} \; \big\{\Delta_\lambda \vdash P_\lambda \approx Q_\lambda : I_\lambda \to O_\lambda\big\}_{\lambda \in \nat}\]
gives us functions
\begin{align*}
\xi & : \nat \to \{0,\ldots,n-1\} \to \nat \\
\psi & : \nat \to \{0,\ldots,n-1\} \to \nat^{|\Sigma_\type|+1} \to \nat
\end{align*}
such that for each $\lambda \in \nat$, we have
\[\Delta_\lambda \vdash \approxeq{P_\lambda}{Q_\lambda}{I_\lambda}{O_\lambda}{\xi_\lambda}{\psi_\lambda} \tag{$\star$}\]
and for each $1 \leq i \leq n$, we have
\begin{align*}
\xi_{(\cdot)}^i & = \mathsf{O}(\poly(\lambda)) \\
\psi_{(\cdot)}^i & = \mathsf{O}\big(\poly(\lambda,t_1,\ldots,t_{|\Sigma_\type|})\big)
\end{align*}
In particular, there are polynomials $p^i_\cnt(\lambda)$ with $N^i_\cnt \in \nat$ such that $\xi_\lambda^i \leq p^i_\cnt(\lambda)$ if $\lambda \geq N^i_\cnt$, and polynomials $p^i_\context(\lambda,t_1,\ldots,t_{|\Sigma_\type|})$ with $N^i_\context \in \nat$ such that
\[\psi_\lambda^i(t_1,\ldots,t_{|\Sigma_\type|}) \leq p^i_\context(\lambda,t_1,\ldots,t_{|\Sigma_\type|})\] if $\lambda \geq N^i_\context$ and $t_1,\ldots,t_{|\Sigma_\type|} \geq N^i_\context$.

Since $\big\{\int{-}_\lambda\big\}_{\lambda \in \nat}$ is PPT, we have a polynomial $C_\sem(\lambda)$, a negligible function $\eta_\sem(\lambda)$, and a $N_\sem \in \nat$ such that:

\begin{itemize}
\item[]
\begin{itemize}
\item \emph{for all type symbols $\type$, $|\type|_\lambda \leq C_\sem(\lambda)$ if $\lambda \geq N_\sem$,}

\item \emph{for all function symbols $\func$, $\int{\func}_\lambda$ is computable by a deterministic Turing Machine $\TM_\lambda$ with symbols $\mathsf{0}, \mathsf{1}$, and both the number of states and the runtime of $\TM_\lambda$ are $\leq C_\sem(\lambda)$ if $\lambda \geq N_\sem$, and}

\item \emph{for all distribution symbols $\dist$, $\int{\dist}_\lambda$ is computable up to an error $\eta_\sem(\lambda)$ by a probabilistic Turing Machine $\TM_\lambda$ with symbols $\mathsf{0}, \mathsf{1}$, and both the number of states and the runtime of $\TM_\lambda$ are $\leq C_\sem(\lambda)$ if $\lambda \geq N_\sem$.}
\end{itemize}
\end{itemize}

\noindent To prove
\[\big\{\int{-}_\lambda\big\}_{\lambda \in \nat} \; \mathlarger{\mathlarger{\vDash}} \; \big\{\Delta_\lambda \vdash P_\lambda \approx Q_\lambda : I_\lambda \to O_\lambda\big\}_{\lambda \in \nat},\]
assume a polynomial $C_\adv(\lambda)$ and a negligible function $\eta_\adv(\lambda)$. Since the ambient asymptotic theory is sound under the ambient family of interpretations, we have the computational indistinguishability assumption
\[\big\{\int{-}_\lambda\big\}_{\lambda \in \nat} \; \mathlarger{\mathlarger{\vDash}} \; \big\{\Delta^i_\lambda \vdash P^i_\lambda \approx Q^i_\lambda : I^i_\lambda \to O^i_\lambda\big\}_{\lambda \in \nat}.\]
Define
\[C^i_\context(\lambda) \coloneqq p^i_\context\big(\lambda,C_\sem(\lambda) + N^i_\context,\ldots,C_\sem(\lambda) + N^i_\context\big),\]
and apply the computational indistinguishability assumption above to the polynomial
\[\mathcal{P}\big(C_\sem(\lambda),C_\adv(\lambda),C^i_\context(\lambda)\big)\]
and the negligible function $\max\big(\eta_\sem(\lambda),\eta_\adv(\lambda)\big)$.

This yields a negligible function $\varepsilon^i(\lambda)$ with an $N^i \in \nat$ such that for any $\lambda \geq N^i$ and any adversary $\Adv^i$ for protocols $\Delta^i_\lambda \vdash I^i_\lambda \to O^i_\lambda$ under the interpretation $\int{-}_\lambda$ with the property that \[|\Adv^i| \leq \mathcal{P}\big(C_\sem(\lambda),C_\adv(\lambda),C^i_\context(\lambda)\big)\] and $\err(\Adv^i) \leq \max\big(\eta_\sem(\lambda),\eta_\adv(\lambda)\big)$, we have
\[\Big|\mathsf{Pr}\big[\interaction{\Adv}{P^i_\lambda}{\int{-}_\lambda} = 1\big] - \mathsf{Pr}\big[\interaction{\Adv}{Q^i_\lambda}{\int{-}_\lambda} = 1\big]\Big| \leq \varepsilon^i(\lambda).\]%

\noindent We can now define our desired negligible function as
\[\varepsilon(\lambda) \coloneqq \sum_{i = 1}^n \xi^i_\lambda * \big(\varepsilon^i(\lambda) + 2 * C^i_\context(\lambda) * \eta_\sem(\lambda)\big)\]
The negligibility of $\varepsilon(\lambda)$ follows easily: if $\lambda \geq N^i_\cnt$, then
\begin{align*}
\xi^i_\lambda * \big(\varepsilon^i(\lambda) + 2 * C^i_\context(\lambda) * \eta_\sem(\lambda)\big)
\leq p^i_\cnt(\lambda) * \big(\varepsilon^i(\lambda) + 2 * C^i_\context(\lambda) * \eta_\sem(\lambda)\big),
\end{align*}
thus it suffices to show that this latter function is negligible. But this is immediate from the negligibility of $\varepsilon^i(\lambda)$, the negligibility of $\eta_\sem(\lambda)$, and the fact that $p^i_\cnt(\lambda)$ and $C^i_\context(\lambda)$ are polynomials. Define
\[N \coloneqq \max\big(N_\sem,N^1_\context,\ldots,N^n_\context,N^1,\ldots,N^n\big).\]
Now assume $\lambda \geq N$ and take any adversary $\Adv$ for protocols $\Delta_\lambda \vdash I_\lambda \to O_\lambda$ under the interpretation $\int{-}_\lambda$, such that $|\Adv| \leq C_\adv(\lambda)$ and $\err(\Adv) \leq \eta_\adv(\lambda)$. We aim to show that
\[\Big|\mathsf{Pr}\big[\interaction{\Adv}{P_\lambda}{\int{-}_\lambda} = 1\big] - \mathsf{Pr}\big[\interaction{\Adv}{Q_\lambda}{\int{-}_\lambda} = 1\big]\Big| \leq \sum_{i = 1}^n \xi^i_\lambda * \big(\varepsilon^i(\lambda) + 2 * C^i_\context(\lambda) * \eta_\sem(\lambda)\big)\]
But this is precisely the conclusion of Theorem~1 applied to the derivation $(\star)$. It thus suffices to prove the hypotheses of Theorem~1. Among these, the only non-trivial assumption is
\[\psi^i_\lambda\big(|\type_1|,\ldots,|\type_{|\Sigma_\mathsf{t}|}|\big) \leq C^i_\context(\lambda).\]
We show this via the following sequence of inequalities:
\begin{align*}
& \psi^i_\lambda\big(|\type_1|_\lambda,\ldots,|\type_{|\Sigma_\mathsf{t}|}|_\lambda\big) \leq \\
& \psi^i_\lambda\big(C_\sem(\lambda),\ldots,C_\sem(\lambda)\big) \leq \\
& \psi^i_\lambda\big(C_\sem(\lambda) + N^i_\context,\ldots,C_\sem(\lambda) + N^i_\context\big) \leq \\
& p^i_\context\big(\lambda,C_\sem(\lambda) + N^i_\context,\ldots,C_\sem(\lambda) + N^i_\context\big) = C^i_\context(\lambda).
\end{align*}
The first inequality follows since the function $\psi^i_\lambda : \nat^{|\Sigma_\type|} \to \nat$ is monotonically increasing in each argument and $|\type_1|_\lambda,\ldots,|\type_{|\Sigma_\type|}|_\lambda \leq C_\sem(\lambda)$ by assumption since $\lambda \geq N_\sem$. The second inequality is again monotonicity of $\psi^i_\lambda$, and the third follows from the definition of $p^i_\context$ since $\lambda \geq N^i_\context$ and $C_\sem(\lambda) + N^i_\context \geq N^i_\context$.
\end{proof}

%%%%%%%%%%%%%%%%%%%%%%%%%%%%%%%%%%%%%%%%%%%%%%%%%%%%%%%%%%%%%%%%%

\section{Absorption: Proof Sketch for Lemma~2}\label{sec:absorption}

\begin{lemma*}[Absorption]
There exists a polynomial $\mathcal{P}(x,y,z) \geq y$ such that for any
\begin{itemize}
\item interpretation $\int{-}$ for $\Sigma$ for which there are $C_\sem \in \nat$, $\eta_\sem \in \rat_{\geq 0}$ such that
\begin{itemize}
\item for all type symbols $\type$, $|\type| \leq C_\sem$,

\item for all function symbols $\func$, $\int{\func}$ is computable by a TM with symbols $\mathsf{0}, \mathsf{1}$ such that the number of states and the runtime are $\leq C_\sem$, and

\item for all distribution symbols $\dist$, $\int{\dist}$ is computable up to error $\eta_\sem$ by a probabilistic TM with symbols $\mathsf{0}, \mathsf{1}$ such that the number of states and the runtime are $\leq C_\sem$,
\end{itemize}

\item adversary $\Adv$ for protocols of type $\Delta \vdash I \to O_1 \cup O_2$ such that $|\Adv| \leq C_\adv$ and $\err(\Adv) \leq \eta_\adv$ for some $C_\adv \in \nat$ and $\eta_\adv \in \rat_{\geq 0}$,

\item protocol $\Delta \vdash Q : I \cup O_1 \to O_2$,
\end{itemize}
we have an adversary $\Adv_\mathcal{R}$ for protocols $\Delta \vdash I \cup O_2 \to O_1$ with
\[|\Adv_\mathcal{R}| \leq \mathcal{P}\big(C_\sem,C_\adv,\tmnorm{Q}(|\type_1|,\ldots,|\type_{|\Sigma_\mathsf{t}|}|)\big)\]
and $\err(\Adv_\mathcal{R}) \leq \max(\eta_\sem,\eta_\adv)$ such that for any protocol $\Delta \vdash P : I \cup O_2 \to O_1$ we have
\[\Big|\mathsf{Pr}\big[\interaction{\Adv}{\Par{P}{Q}}{\int{-}} = 1\big] - \mathsf{Pr}\big[\interaction{\Adv_\mathcal{R}}{P}{\int{-}} = 1\big]\Big| \leq \tmnorm{Q}(|\type_1|,\ldots,|\type_{|\Sigma_\mathsf{t}|}|) * \eta_\sem.\]
\end{lemma*}

\begin{proof}[Sketch]
Let $\mathsf{context} \coloneqq \tmnorm{Q}(|\type_1|,\ldots,|\type_{|\Sigma_\mathsf{t}|}|)$, and let $O_1^\mathsf{min}$ be the minimal set of query inputs to $Q$ from among $O_1$. In other words, $O_1^\mathsf{min}$ contains precisely those channels of $O_1$ that $Q$ reads from. The reason for replacing $O_1$ with $O_1^\mathsf{min}$ is that we do not have a bound on the size of the former, but the size of the latter is bounded by the number of occurrences of the query annotation \textsf{``$\mathsf{input}$-$\mathsf{to}$-$\mathsf{query}$''}, and this is in turn bounded by $\mathsf{context}$. Let 
\[\Adv \coloneqq \big(\Delta', I', O', \phi, \#_\round, \#_\tape, \Symb, \St, s_\star, \Step, \big\{\In_o\big\}_{o \, \in \, I'}, \big\{\Out_i\big\}_{i \, \in \, O'}, \Dec\big)\]
be the adversary for protocols of type $\Delta \vdash I \to O_1 \cup O_2$. Define the reduced adversary $\Adv_\mathcal{R}$ for protocols of type $\Delta \vdash I \cup O_2 \to O_1$ as follows:
\[\Adv_\mathcal{R} \coloneqq \big(\Delta', I'_\mathcal{R}, O'_\mathcal{R}, \phi, \#_\round^\mathcal{R}, \#_\tape^\mathcal{R}, \Symb^\mathcal{R}, \St^\mathcal{R}, s^\mathcal{R}_\star, \Step^\mathcal{R}, \big\{\In^\mathcal{R}_o\big\}_{o \, \in \, I'_\mathcal{R}}, \big\{\Out^\mathcal{R}_i\big\}_{i \, \in \, O'_\mathcal{R}}, \Dec^\mathcal{R}\big)\]
Here:
\begin{itemize}
\item the set of \emph{inputs} is $I'_\mathcal{R} \coloneqq (I' \, \cup \, \phi^\star(O_1)) \cup \phi^\star(O_1^\textsf{min})$;
\item the set of \emph{outputs} is $O'_\mathcal{R} \coloneqq O' \cup \phi^\star(O_2)$;
\item the number of \emph{rounds} is $\#_\round^\mathcal{R} \coloneqq \#_\round * \mathsf{context} + \mathsf{context}^2 + 2 * \#_\round + \mathsf{context} + 1$;
\item the number of \emph{tapes} is $\#_\tape^\mathcal{R} \coloneqq \#_\tape + 1$;
\item the set of \emph{symbols} is
\[\Symb^\mathcal{R} \coloneqq \big\{r \; | \; 0 \leq r \leq \#_\round\big\} \, \bigsqcup \ \textsf{ProtEncSymb} \bigsqcup \{\textsf{``$\rightleftharpoons$''}, \textsf{``$\#$''}\} \, \bigsqcup \, \textsf{Symb},\]
where $\mathsf{ProtEncSymb}$ is the disjoint union of the following sets: $\{\textsf{`` ''}\}$, $\mathsf{Punc}$, $\mathsf{KeyWords}$,
the set $\Sigma_\func$ of function symbols declared in $\Sigma$, the set $\Sigma_\dist$ of distribution symbols declared in $\Sigma$, the set \[\big\{n \;|\; 0 \leq n < \textit{variable-index bound}(Q)\big\},\] and the set
\begin{align*}
& \big\{n \;|\; 0 \leq n < \textit{channel-index bound}(Q)\big\} \, \bigcup \\ & \big\{m + n \; | \; m \in \phi^\star(I \cup O_1^\mathsf{min} \cup O_2) \textit{ and } 0 \leq n \leq \textit{channel-index bound}(Q)\big\};
\end{align*}
\item the set of \emph{states} is
\[\St^\mathcal{R} \coloneqq \big\{\textsf{``}(\textsf{''} + r + b + \mathsf{prot} + \textsf{``}\rightleftharpoons\textsf{''} + \mathsf{adv} + \textsf{``})\textsf{''}\big\},\]
where
\begin{itemize}
\item the round counter $0 \leq r \leq \#_\round$ denotes the number of rounds remaining,
\item the Boolean $b \in \{0,1\}$ indicates whether we are processing the adversary code ($0$) or the absorbed protocol code ($1$),
\item $\mathsf{adv} \in \St$ is the original adversary code, and
\item $\mathsf{prot}$ is the absorbed protocol code,
\end{itemize}
\item the \emph{initial state} $s^\mathcal{R}_\star$ sets:
\begin{itemize}
\item $r \coloneqq \#_\round$,
\item $b \coloneqq 1$,
\item $\mathsf{adv} \coloneqq s_\star$, and
\item $\mathsf{prot}$ is the encoding of the protocol $Q$ with every channel read from $O_1^\textsf{min}$ annotated with \textsf{``$\mathsf{input}$-$\mathsf{to}$-$\mathsf{query}$''}.
\end{itemize}
\end{itemize}
The TM $\Step^\mathcal{R}$ executes the encoded protocol code by first searching for a channel read annotated with \textsf{``$\mathsf{input}$-$\mathsf{to}$-$\mathsf{query}$''}; if it finds one, it updates the annotation to \textsf{``$\mathsf{input}$-$\mathsf{queried}$''} and performs the query. Otherwise all inputs from $O_1^\textsf{min}$ have been queried. We thus enter a second phase, where we search for a reaction that computes; if we find one, we perform the computation. Otherwise the protocol computation terminates and we set the bit $b$ to $0$. Our proof yields the following polynomial:
\begin{align*}
P(x,y,z) & \coloneqq y^2 + 8yz + 15z^2 + (|\Sigma_f| + |\Sigma_d| + 1)x + 34y + 47z \, + \\ & \hspace{20pt} (|\mathsf{Punc}| + |\mathsf{KeyWords}| + |\Sigma_f| + |\Sigma_d| + 161)
\end{align*}
\end{proof}

\section*{Acknowledgments}

This work is partially supported by the NGI0 Core Fund (\url{https://nlnet.nl/core}), a fund established by NLnet (\url{https://nlnet.nl}) with financial support from the European Commission's Next Generation Internet programme (\url{https://ngi.eu}), under the aegis of DG Communications Networks, Content and Technology under grant agreement Nr. 101092990. 

\end{document}